\documentclass{WileySTAT-arxiv} 

\usepackage[T1]{fontenc}
\usepackage[utf8]{inputenc}

\usepackage[bookmarksopen,bookmarksnumbered,citecolor=blue,urlcolor=blue]{hyperref} 
\usepackage{natbib}     
\usepackage{bm}
\usepackage{dcolumn}    
\usepackage{siunitx}
\usepackage{subcaption}
\newcolumntype{d}[1]{D{.}{.}{#1}}

\runninghead{Qirui Fu, Yupeng Jiang*, Minchen Li*}{An Implicit CK-MPM for Computational Solid Mechanics}

\theoremstyle{plain}

\begin{document}

\title{An Implicit Compact-Kernel Material Point Method for Computational Solid Mechanics}

\author{Qirui Fu\affil{a},
        Yupeng Jiang\affil{b}\corrauth, Minchen Li\affil{a,c}\corrauth}

\address{%
\affilnum{a}Computer Science Department, Carnegie Mellon University \\
\affilnum{b}IBNM, Leibniz Universität Hannover \\
\affilnum{c}Genesis AI}

\articlenote{
Yupeng Jiang and Minchen Li jointly supervised this work.
}
\corremail{yupeng.jiang@ibnm.uni-hannover.de, minchernl@gmail.com}

\begin{abstract}
The numerical performance of the material point method (MPM) is strongly governed by the particle-grid kernel, which controls the trade-off among smoothness, locality, numerical diffusion, contact accuracy, and computational cost. Although wide-support smooth kernels can effectively suppress cell-crossing instability, they often introduce increased numerical diffusion, artificial contact gaps, and higher transfer cost. In contrast, the suitability of compact-kernel designs for implicit computational solid mechanics remains unclear. In this work, we develop an implicit formulation of the Compact-Kernel Material Point Method (CK-MPM) building on the framework of \citet{liu2025ck}, and assess its performance through benchmark problems in linear and nonlinear solid mechanics. Numerical assessment comprises cantilever bending, Hertzian contact, narrow-clearance free fall, and colliding hyperelastic rings, supplemented by a transfer-only diffusion test and spatial and temporal refinement studies. Implicit CK-MPM mitigates the cell-crossing stress noise and numerical energy loss observed in linear MPM, while retaining bending and energy responses comparable to those of quadratic B-spline MPM. Meanwhile, its more localized transfers improve contact-pressure accuracy and reduce the artificial contact gaps and early-contact artifacts associated with the wider-support quadratic kernel. These results demonstrate that the developed implicit formulation retains CK-MPM’s favorable balance between smoothness and transfer locality in static and dynamic solid-mechanics analysis. 
\end{abstract}

\keywords{Material point method; Compact-kernel; Implicit time integration; Large deformation; Contact mechanics}

\maketitle

\section{Introduction}\label{sec1}

The material point method (MPM) has become an increasingly important numerical approach in computational continuum mechanics (\citet{bardenhagen2000material,de2020material}). Combining a hybrid Lagrangian–Eulerian description with particle-based discretization, MPM offers distinct advantages for problems involving large deformations (\citet{wikeckowski2004material,jiang2016material,zhao2024mapped}), complex constitutive responses (\citet{solowski2015evaluation, askari2016intrusion,baumgarten2019general,jiang2022hybrid}), and severe kinematic motions (\citet{ma2009comparison}). These features make it particularly attractive for simulations in which conventional mesh-based methods may suffer from excessive mesh distortion or difficulties in tracking evolving material interfaces (\citet{de2020material}). As a result, MPM has been successfully applied to a broad spectrum of problems, including physics-based animation (\citet{wolper2019cd, wolper2020anisompm,li2024dynamic,li2022energetically,wang2020massively}), virtual reality \citep{luo2025vr}, topology optimization \citep{li2021lagrangian}, geophysical flows (\citet{jiang2020hybrid,jiang2023erosion,liang2023revealing,zhao2023coupled,jiang2024impact}), and dynamic fracture of solids (\citet{zeng2023explicit}), where it has delivered robust predictive capability and valuable physical insight.

At the core of the material point method lies the choice of shape function, or kernel function, which governs the transfer of material information between Lagrangian material points and Eulerian grid nodes. In each computational loop, state variables carried by the material points are first projected to the background grid through the particle-to-grid (P2G) mapping defined by the kernel functions (\citet{bardenhagen2004generalized,jiang2016material}). The governing equations are then solved on the grid, after which the updated nodal quantities are transferred back to the material points through the corresponding grid-to-particle (G2P) mapping for the subsequent update of material states. Since these bidirectional transfers are repeated throughout the computation, the kernel function plays a central role in controlling the accuracy, stability, and overall numerical performance of MPM.

In its original form, MPM employs linear shape functions that are only $C^0$-continuous, with discontinuous first derivatives across cell boundaries (\citet{bardenhagen2000material}). As a result, when material points cross grid cells, the abrupt change in the kernel gradient may induce pronounced numerical artifacts, commonly referred to as cell-crossing noise. To alleviate this deficiency, one class of methods reinterprets material points as finite-sized continuum domains, thereby regularizing the transfer process during boundary crossing. Representative developments along this line include the Generalized Interpolation Material Point Method (GIMP) (\citet{bardenhagen2004generalized}) and the Convected Particle Domain Interpolation (CPDI) method (\citet{sadeghirad2011convected,sadeghirad2013second}). Another widely adopted strategy is to employ smoother higher-order kernels, such as quadratic or cubic B-splines, which provide improved continuity across cell boundaries and thus significantly enhance numerical stability. They are often used together with the Affine particle-in-cell (APIC) scheme (\citet{jiang2017angular}) or Moving Least Squares MPM (MLS-MPM) (\citet{hu2018moving,cao2025unstructured}). Collectively, these developments have greatly reduced cell-crossing artifacts and broadened the applicability of MPM to challenging large-deformation problems.

A major drawback of these numerically enhanced kernels is that their improved smoothness is typically achieved at the expense of enlarged support, such that a larger number of neighboring grid nodes participate in the P2G and G2P transfers (\citet{sadeghirad2013second,gan2018enhancement,yamaguchi2021extended, li2024contact}). This wider stencil increases the computational work associated with particle--grid transfer because more grid nodes participate in each P2G and G2P operation. More importantly, the enlarged support reduces transfer locality and tends to amplify numerical diffusion (\citet{jiang2017angular}), thereby smearing sharp kinematic features and material interfaces. In problems involving separation and fracture, such excessive smoothing may even manifest as spurious numerical cohesion or adhesion (\citet{wolper2019cd,liang2024mortar,bonus2025validating}). In contact simulations, wide-support kernels may further induce artificial contact gaps, which distort the contact evolution and compromise the accuracy of contact force transmission (\citet{chen2023enriched, kakouris2025extended}). Existing remedies often rely on additional geometric descriptions and explicit boundary-tracking procedures (\citet{bing2019b,liu2020ils,liang2024mortar,zhang2025nitsche}), but these enhancements come with considerable algorithmic and implementation complexity. Their use is particularly challenging in problems involving multi-body interaction, large rotation, separation, and evolving fracture, where material boundaries must be continuously identified and updated. These limitations strongly motivate the development of an MPM formulation that can retain sufficient kernel smoothness while preserving compact support and transfer locality.

To address these requirements, the Compact-Kernel Material Point Method (CK-MPM) provides a promising direction (\citet{liu2025ck}). Originally developed in an explicit setting for graphics-oriented simulations, CK-MPM was designed to reduce artificial numerical gaps while retaining a compact transfer stencil. Its key feature is a specially constructed dual-grid framework that enables particle-grid transfer to remain confined to the containing cell while preserving $C^2$ continuity through a specially designed function. As a result, CK-MPM offers an attractive balance between kernel smoothness and transfer locality, while limiting the number of grid nodes involved in the transfer of each material point. This compact transfer stencil is also computationally attractive because it limits the particle--grid interactions required by each transfer.

Despite these appealing features, the applicability of CK-MPM to computational mechanics has not yet been systematically investigated, and its algorithmic performance in an implicit setting remains unclear. This constitutes a nontrivial gap in current knowledge. In contrast to explicit time integration, the implicit equilibrium problem for CK-MPM must account for the coupling induced by its dual-grid interpolation: each particle’s deformation depends on nodal quantities from both grids, and this dependence must be represented consistently in the residual and tangent. The central question addressed here is therefore how to formulate and solve the coupled implicit system while preserving the original transfer construction, and how the resulting method performs in stiffness-dominated deformation, contact, and large-deformation dynamics. Such issues are particularly important in computational mechanics applications, where implicit MPM is often preferred because of its ability to accommodate large time steps and to provide robust treatment of stiff problems. 

In this work, we develop an implicit formulation of CK-MPM for static and dynamic solid mechanics. Retaining the compact-kernel discretization introduced by Liu et al. (2025), we formulate the backward-Euler grid update as an incremental-potential minimization and derive the associated residual and consistent tangent. The resulting nonlinear system is solved using a projected Newton method. A corresponding displacement-based formulation is developed for direct solution of static equilibrium. We further systematically assess its performance through a set of benchmark problems representative of both linear and nonlinear solid mechanics. Our objective is to establish, through rigorous numerical evidence, that the compact-kernel philosophy of CK-MPM remains advantageous in a computational mechanics setting beyond its original explicit, graphics-oriented context. To this end, we consider a diverse set of benchmark tests, including gravity-induced cantilever bending, circle-to-plane Hertzian contact, interface diffusion, and colliding ring kinematics, which collectively examine the method in terms of accuracy, robustness, transfer locality, and contact fidelity under an implicit framework. The results show that implicit CK-MPM preserves the distinctive strengths of the compact-kernel design, reducing transfer-induced diffusion and artificial contact effects while maintaining smooth and robust behavior across a range of solid-mechanics problems.

\section{Fundamentals of CK-MPM}

\subsection{Strong Form and Weak Form}
In this work, we consider a material domain $\Omega\in\mathbb{R}^d$, where $d\in\{1,2,3\}$ and whose boundary is denoted by $\partial\Omega$. With thermal effects neglected, the conservation of momentum equation can be written as:
\begin{equation}
        \rho \ddot{\bm{u}} = \nabla\cdot\bm{\sigma} + \rho\bm{b}, 
\end{equation}
where $\rho$ is the mass density of the material, $\bm{u}$ is the displacement field, $\bm{\sigma}$ is the Cauchy stress tensor, and $\bm{b}$ is the body force per unit mass. The Dirichlet and Neumann boundary conditions are specified as:
\begin{equation}
        \bm{u} = \bar{\bm{u}} \ \text{on}  \ \partial\Omega_{D}, \quad
        \bm{\sigma}\bm{n} = \bar{\bm{t}} \ \text{on} \ \partial\Omega_{N},
\end{equation}
where $\bm{n}$ denotes the outward unit normal to the boundary and
$\bar{\bm{t}}$ is the prescribed traction.

Let $\bm{w}$ be the test function vanishing on $\partial\Omega_{D}$. Multiplying the strong form by $\bm{w}$ and integrating over $\Omega$ gives
\begin{equation}
        \int_\Omega \bm{w}\cdot(\rho\ddot{\bm{u}} - \nabla\cdot\bm{\sigma} - \rho\bm{b})\, \mathrm{d}\Omega = 0.
\end{equation}
Applying integration by parts gives:
\begin{equation}
        \int_\Omega \rho\bm{w}\cdot\ddot{\bm{u}}\, \mathrm{d}\Omega + \int_\Omega \nabla\bm{w}:\bm{\sigma}\, \mathrm{d}\Omega = 
        \int_\Omega \rho\bm{w}\cdot\bm{b}\, \mathrm{d}\Omega + \int_{\partial\Omega_{N}} \bm{w}\cdot\bar{\bm{t}}\, \mathrm{d}(\partial\Omega_{N}).
\end{equation}
This is the standard weak form of the momentum equation. The first term on the left-hand side represents the inertial contribution, while the second term corresponds to the internal force arising from the stress field. The two terms on the right-hand side represent external forces, including body forces and traction forces prescribed on the Neumann boundary.

\subsection{MPM Discretization} \label{sec:mpm_discretization}
In Material Point Method (MPM) \citep{sulsky1995application}, the continuum domain is decomposed into a collection of particle domains $\Omega_{p}$:
\begin{equation}
        \Omega = \bigcup_p \Omega_p.
\end{equation}
With this discretization, the volume integral can be expressed as:
\begin{equation}
        \int_\Omega f(\bm{x})\, \mathrm{d}\Omega = \sum_p \int_{\Omega_{p}} f(\bm{x})\, \mathrm{d}\Omega.
\end{equation}
Within each particle domain, standard MPM assumes all physical quantities are constant, which means: $\rho(\bm{x}) = \rho_p$, $\ddot{\bm{u}}(\bm{x}) = \ddot{\bm{u}}_p$, and $\sigma(\bm{x}) = \sigma_p$.

Meanwhile, the test function is discretized using grid shape functions $N_i(\bm{x})$:
\begin{equation}
        \bm{w}(\bm{x}) = \sum_i \bm{w}_i N_i(\bm{x}),
\end{equation}
where $\bm{w}_i$ is a nodal test value. Substituting the particle-domain and grid-shape-function discretizations, the inertia term in the weak form becomes:
\begin{align}
        \int_\Omega \rho\bm{w}\cdot\ddot{\bm{u}}\, \mathrm{d}\Omega &= \sum_p \rho_p\ddot{\bm{u}}_p\int_{\Omega_{p}} \bm{w}(\bm{x})\, \mathrm{d}\Omega \\
        &= \sum_p\rho_p\ddot{\bm{u}}_p\ \left(\sum_i\bm{w}_i \int_{\Omega_{p}} N_i(\bm{x})\, \mathrm{d}\Omega \right) \\
        &= \sum_i\bm{w}_i\ \left(\sum_p\rho_p\ddot{\bm{u}}_p \int_{\Omega_{p}} N_i(\bm{x})\, \mathrm{d}\Omega \right).
\end{align}
Similarly, the internal force term becomes:
\begin{align}
\int_\Omega \nabla \bm{w} : \bm{\sigma} \, \mathrm{d}\Omega
&=
\sum_p
\int_{\Omega_p}
\nabla \bm{w} : \bm{\sigma}
\, \mathrm{d}\Omega
\\
&=
\sum_p
\int_{\Omega_p}
\left(
\sum_i \bm{w}_i \nabla N_i(\bm{x})
\right)
: \bm{\sigma}_p
\, \mathrm{d}\Omega
\\
&=
\sum_i \bm{w}_i
\left( \sum_p
\int_{\Omega_p}
\nabla N_i(\bm{x})
: \bm{\sigma}_p
\, \mathrm{d}\Omega
\right)
\\
&=
\sum_i \bm{w}_i
\left( \sum_p \bm{\sigma}_p
\int_{\Omega_p}
\nabla N_i(\bm{x})
\, \mathrm{d}\Omega
\right).
\end{align}
Since $\bm{w}_i$ are arbitrary, we choose it successively as the unit vector for each nodal degree of freedom. Substituting this choice into the weak form yields, for each node $i$:
\begin{equation}
        \sum_p\rho_p\ddot{\bm{u}}_p \int_{\Omega_{p}} N_i(\bm{x})\, \mathrm{d}\Omega
        = -\sum_p \bm{\sigma}_p \int_{\Omega_p} \nabla N_i(\bm{x})\, \mathrm{d}\Omega + \bm{f}_i^{ext}.
\end{equation}
Introducing particle mass $m_p$, particle density becomes $\rho_p = \frac{m_p}{V_p}$. Meanwhile, integration of the grid shape function and its gradient have first-order estimations:
\begin{align}
        \int_{\Omega_{p}} N_i(\bm{x})\, \mathrm{d}\Omega \approx V_p N_i(\bm{x}_p), \\
        \int_{\Omega_{p}} \nabla N_i(\bm{x})\, \mathrm{d}\Omega \approx V_p \nabla N_i(\bm{x}_p).
\end{align}
Together, we have
\begin{equation}
        \sum_p m_p\ddot{\bm{u}}_p N_i(\bm{x}_p)
        = -\sum_p V_p\bm{\sigma}_p \nabla N_i(\bm{x}_p) + \bm{f}_i^{ext}.
\end{equation}
Because $N_i(\bm{x}_p)$ is determined by grid node $i$ and particle $p$, we define the transfer weight $\omega_{ip} = N_i(\bm{x}_p)$ to represent the associativity between the grid and particles. Thus, our governing equation becomes:
\begin{equation}
        \sum_p m_p\ddot{\bm{u}}_p \omega_{ip}
        = -\sum_p V_p\bm{\sigma}_p \nabla \omega_{ip} + \bm{f}_i^{ext}.
\label{eq:dis_weak_form}
\end{equation}

To solve this discretized weak form, the MPM framework consists of four steps:
\begin{itemize}
        \item \textbf{Particle-to-Grid (P2G) Transfer:} transfer mass and momentum from particle to grid
        \item \textbf{Grid Update:} calculate force on the grid node, and perform nodal momentum update
        \item \textbf{Grid-to-Particle (G2P) Transfer:} transfer velocity from grid to particle and update particle deformation gradient
        \item \textbf{Particle Advection:} update particle position
\end{itemize}

\subsection{Kernel Function and Basis Selection}
\label{sec:ckmpm_kernel}
From Eq.~\eqref{eq:dis_weak_form}, the choice of grid basis function $N_i(\bm{x})$ directly determines the smoothness, consistency, and stability properties of the method. Many MPM variants can be interpreted as using different grid basis functions. In most cases, we define the one-dimensional kernel function $\mathcal{K}(x)$ to satisfy:
\begin{itemize}
        \item $\mathcal{K}(x) = \mathcal{K}(-x)$
        \item $\int_{-\infty}^{+\infty}\mathcal{K}(x)\, \mathrm{d}x = 1$
        \item $\mathcal{K}(x)\geq 0$
\end{itemize}
and define $N_i(\bm{x})$ as its tensor product:
\begin{equation}
        N_i(\bm{x}) = \prod_{d=1}^{dim} \mathcal{K}(\frac{x_{i, d} - x_{d}}{h}),
\end{equation}
where $x_{i, d}, x_d$ are the d-th component of positions and $h$ represents the grid spacing.

A straightforward choice of kernel function is the piecewise linear function:
\begin{equation*}
\mathcal{K}(d) =
        \begin{cases}
                1 - |d|, & -1 \leq d \leq 1 \\
                0, & \text{otherwise}.
        \end{cases}
\end{equation*}
However, this function is only $C^0$-continuous at $x=-1, 0, 1$, which can result in discontinuous forces when particles cross cell boundaries, potentially leading to numerical instability. Thus, to avoid this issue, many MPM frameworks use the $C^1$-continuous quadratic B-spline kernel function \citep{steffen2008analysis}
\begin{equation*}
\mathcal{K}(d) =
        \begin{cases}
                \frac{1}{2}d^2 + \frac{3}{2}d + \frac{9}{8}, & -\frac{3}{2} \leq d < -\frac{1}{2} \\
                -d^2 + \frac{3}{4}, & -\frac{1}{2} \leq d < \frac{1}{2} \\
                \frac{1}{2}d^2 - \frac{3}{2}d + \frac{9}{8}, & \frac{1}{2} \leq d \leq \frac{3}{2} \\
                0, & \text{otherwise}.
        \end{cases}
\end{equation*}
The support radius of this kernel is 1.5, compared with 1 in linear kernel. With this kernel function, each material point in MPM is associated with $9$ grid nodes in $2D$ and $27$ in $3D$. 

\begin{figure}[htbp]
        \centering
        \includegraphics[width = 0.5\textwidth]{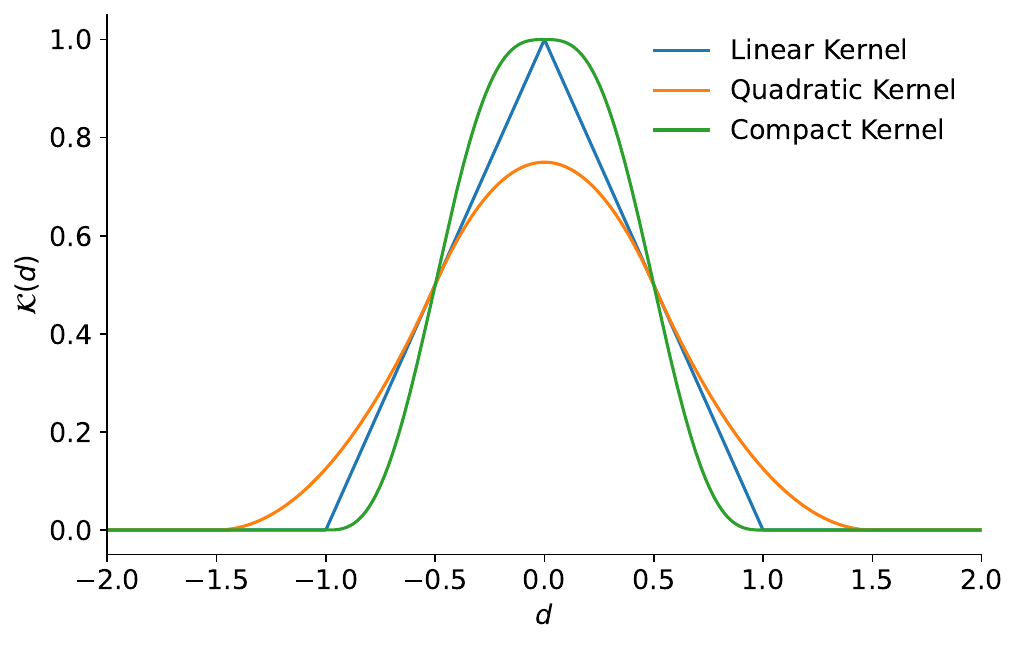}
        \caption{Comparison of the three kernel functions used in this study. The compact kernel combines the principal advantages of the linear and quadratic kernels, possessing a support radius of 1 and being differentiable everywhere.}
        \label{fig:kernel-curve}
\end{figure}

In this study, we use a recently proposed compact kernel from \citet{liu2025ck} whose support radius remains 1 but is sufficiently smooth to avoid cell-crossing instability. The kernel is defined as:
\begin{equation*}
\mathcal{K}(d) = 
        \begin{cases}
                1 - |d| + \frac{1}{2\pi}\sin(2\pi|d|), & -1 \leq d \leq 1 \\
                0, & \text{otherwise}.
        \end{cases}
\end{equation*}
MPM employing this kernel is referred to as CK-MPM, or the compact-kernel material point method. \autoref{fig:kernel-curve} compares the profiles of the linear, quadratic B-spline, and compact kernels, highlighting their differences in support radius and smoothness.

\begin{figure}[htbp]
    \centering

    \begin{subfigure}{0.25\linewidth}
        \centering
        \includegraphics[width=\linewidth]{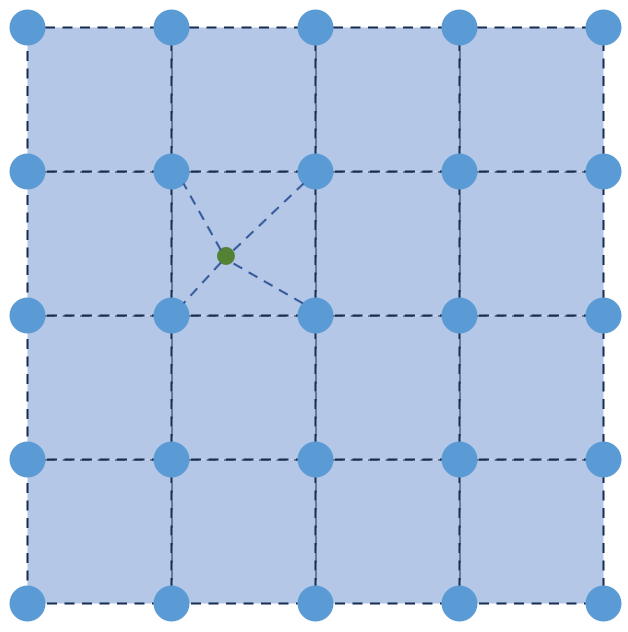}
        \caption{Linear Kernel}
    \end{subfigure}
    \hspace{0.06\linewidth}
    \begin{subfigure}{0.25\linewidth}
        \centering
        \includegraphics[width=\linewidth]{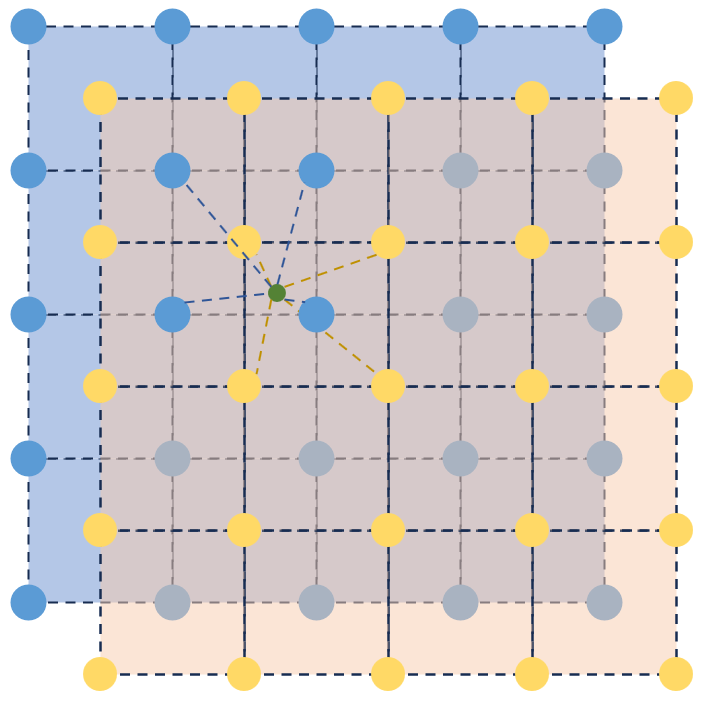}
        \caption{Compact Kernel}
    \end{subfigure}
    \hspace{0.06\linewidth}
    \begin{subfigure}{0.25\linewidth}
        \centering
        \includegraphics[width=\linewidth]{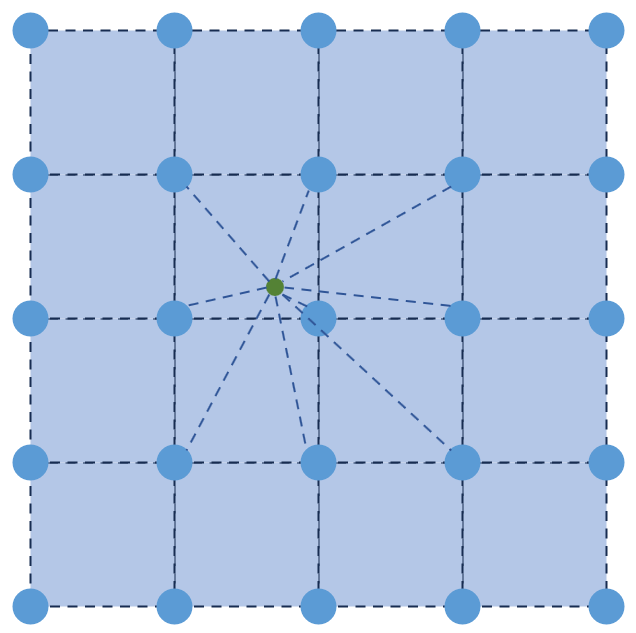}
        \caption{Quadratic Kernel}
    \end{subfigure}
    \caption{Particle–grid connectivity for the three kernel functions. In two dimensions, each particle interacts with 4 grid nodes for the linear kernel and 9 grid nodes for the quadratic kernel. For the compact kernel, each particle interacts with 4 nodes per grid, and two staggered grids are required to satisfy \autoref{eq:recovery_ck}.}
    \label{fig:kernel-radius}
\end{figure}

In MPM, the shape function needs to satisfy two properties at every material point to ensure 1st-order accurate interpolation:
\begin{align}
        \sum_i \omega_{ip} & = 1, \\
        \sum_i \bm{x}_i\omega_{ip} & = \bm{x}_p.\label{eq:recovery} 
\end{align}
It is easy to verify that these two equations hold in quadratic kernel, but Eq.~\eqref{eq:recovery} does not hold for this newly introduced compact kernel, which means utilizing the compact kernel directly would cause numerical issues such as angular momentum loss. Thus, \citet{liu2025ck} introduced a dual-grid system, constructing two staggered uniform grids $\{\mathcal{G}_{-1}, \mathcal{G}_{+1}\}$, where they have an offset of $\frac{1}{2}\Delta x$ in each axis, as shown in \autoref{fig:kernel-radius}. Considering contributions from both grids, the two properties can be satisfied as:
\begin{equation}
        \frac{1}{2}\sum_{k=\pm 1}\sum_{i\in\mathcal{G}_k} \omega_{ip} = 1,
\end{equation}
\begin{equation}
        \frac{1}{2}\sum_{k=\pm 1}\sum_{i\in\mathcal{G}_k} \bm{x}_i\omega_{ip} = \bm{x}_p.
\label{eq:recovery_ck}
\end{equation}

Based on this construction, CK-MPM advances the system by transferring particle information to the two grids, updating the momentum on both grids, and averaging their contributions during grid-to-particle transfer. Even with the dual-grid system, each particle is associated with only 8 grid nodes in 2D and 16 in 3D, fewer than with quadratic kernels.

\section{Method}\label{sec3}

Our method follows the standard MPM pipeline outlined in Section~\ref{sec:mpm_discretization}. In this study, we use APIC \citep{jiang2017angular} to transfer information between grid and material points, leveraging its angular momentum conservation property. For grid update, we perform implicit time integration for enhanced numerical stability. We first reinterpret the dual-grid construction of CK-MPM as a single composite grid with normalized shape functions. Under this representation, the particle-grid transfer, grid momentum equation, and incremental potential take the same form as in standard MPM.

\subsection{Unified Compact-Kernel Discretization}\label{sec3.1}

As introduced in Section~\ref{sec:ckmpm_kernel}, CK-MPM employs two staggered uniform grids $\{\mathcal{G}_{-1},\mathcal{G}_{+1}\}$ to satisfy the linear-reproduction condition in Eq.~\eqref{eq:recovery_ck}. Instead of treating the two grids separately and averaging their contributions during G2P, we regard their nodes as a single composite grid:
\begin{equation}
        \mathcal{G} = \mathcal{G}_{-1}\cup\mathcal{G}_{+1}.
        \label{eq:composite_grid}
\end{equation}
For each node $i\in\mathcal{G}$, we define a normalized shape function as
\begin{equation}
        N_i(\bm{x}) = \frac{1}{2}\prod_{d=1}^{dim}\mathcal{K}\left(\frac{x_{i,d}-x_d}{h}\right),
        \qquad N_{ip}=N_i(\bm{x}_p)=\frac{1}{2}\omega_{ip},
        \label{eq:normalized_ck_kernel}
\end{equation}
where $\mathcal{K}$ is the one-dimensional compact kernel introduced in the preceding section, $h$ is the grid spacing of each staggered grid, and $\omega_{ip}$ denotes the unnormalized compact-kernel weight used in the original dual-grid formulation. The factor $1/2$ is included in the shape function itself; hence, contributions from all nodes are summed over $\mathcal{G}$ without any additional averaging. In particular, $\nabla N_{ip}=\frac{1}{2}\nabla\omega_{ip}$.

From the dual-grid properties established in Eq.~\eqref{eq:recovery_ck}, the normalized shape functions satisfy
\begin{align}
        \sum_{i\in\mathcal{G}}N_{ip} &= 1, \\
        \sum_{i\in\mathcal{G}}\bm{x}_i N_{ip} &= \bm{x}_p.
        \label{eq:normalized_recovery}
\end{align}
Their spatial derivatives consequently satisfy
\begin{equation}
        \sum_{i\in\mathcal{G}}\nabla N_{ip}=\bm{0},\qquad
        \sum_{i\in\mathcal{G}}\bm{x}_i(\nabla N_{ip})^T=\bm{I}.
        \label{eq:normalized_gradient_recovery}
\end{equation}
Thus, the normalized compact kernel satisfies partition of unity and first-order consistency on the composite grid. Although the node distribution differs from a single Cartesian grid, it is the specific interleaved lattice obtained from the two staggered uniform grids, rather than an arbitrary unstructured mesh. Each particle is associated with only $8$ nodes in 2D and $16$ in 3D, as in the original CK-MPM formulation.

This construction is an equivalent representation of the original CK-MPM rather than a change to its particle-grid stencil. In particular, the nodal masses, momenta, and forces are all scaled by $1/2$ compared with the quantities assigned independently to each grid in the original formulation, while the nodal velocities and accelerations remain unchanged when the same scaling is applied consistently. Moreover, the total grid mass now equals the total particle mass, $\sum_{i\in\mathcal{G}}m_i=\sum_p m_p$. This normalized representation allows us to use the conventional MPM weak form and the standard incremental potential without additional dual-grid scaling factors.

\subsection{Particle-Grid Transfer and Particle Advection}\label{sec3.2}

For the composite grid $\mathcal{G}$, in the P2G step, we perform
\begin{equation}
        m_i^n=\sum_p N_{ip}m_p^n
        =\frac{1}{2}\sum_p\omega_{ip}m_p^n,
\end{equation}
\begin{equation}
        m_i^n\bm{v}_i^n
        =\sum_p N_{ip}m_p^n\left(\bm{v}_p^n+\bm{B}_p^n(\bm{D}_p^n)^{-1}(\bm{x}_i-\bm{x}_p^n)\right),
\end{equation}
where $m_i$ and $\bm{v}_i$ are mass and velocity of grid node $i$, superscript $n$ denotes the time step index, and all nodal indices belong to the composite grid $\mathcal{G}$. The matrices $\bm{B}$ and $\bm{D}$ store information of the local affine velocity field. $\bm{D}$ can be calculated as
\begin{equation}
        \bm{D}_p^n=\sum_{i\in\mathcal{G}}N_{ip}(\bm{x}_i-\bm{x}_p^n)(\bm{x}_i-\bm{x}_p^n)^T,
\end{equation}
and $\bm{B}$ used in the next time step is computed during the G2P step as
\begin{equation}
        \bm{B}_p^{n+1}=\sum_{i\in\mathcal{G}}N_{ip}\hat{\bm{v}}_i^{n+1}(\bm{x}_i-\bm{x}_p^n)^T,
\end{equation}
where $\hat{\bm{v}}^{n+1}$ is the grid velocity after the grid update detailed in the next subsection. Because $N_{ip}=\omega_{ip}/2$, these expressions reproduce the original dual-grid APIC transfers when both grids are combined.

During G2P, we additionally compute the velocity and its gradient at particles for the next time step:
\begin{equation}
        \bm{v}_p^{n+1}=\sum_{i\in\mathcal{G}}N_{ip}\hat{\bm{v}}_i^{n+1},
\end{equation}
\begin{equation}
        \nabla\bm{v}_p^{n+1}=\sum_{i\in\mathcal{G}}\hat{\bm{v}}_i^{n+1}(\nabla N_{ip})^T.
        \label{eq:velocity-gradient}
\end{equation}
Then, particle deformation gradient is updated as
\begin{equation}
        \bm{F}_p^{n+1}=(\bm{I}+\Delta t\nabla\bm{v}_p^{n+1})\bm{F}_p^n.
\end{equation}
In the advection step, particle positions are updated as
\begin{equation}
        \bm{x}_p^{n+1}=\bm{x}_p^n+\Delta t\bm{v}_p^{n+1}.
\end{equation}
All interpolation weights and their gradients are evaluated at $\bm{x}_p^n$ and held fixed during the implicit grid solve. Particle positions and the interpolation weights are updated only after the grid solve and G2P step.

\subsection{Grid Update with Implicit Time Integration}\label{sec3.3}
In the grid update step, we perform first-order time integration
\begin{equation}
        \hat{\bm{v}}_i^{n+1}=\bm{v}_i^n+\frac{\bm{f}_i^{n+\lambda}}{m_i^n}\Delta t,\quad 0\leq\lambda\leq 1,
        \label{eq:integrator}
\end{equation}
where $\bm{f}_i^{n+\lambda}=(\bm{f}_i^{int})^{n+\lambda}+(\bm{f}_i^{ext})^{n+\lambda}$ is the nodal force on node $i$, and we use backward Euler by setting $\lambda=1$. Then the momentum equation (Eq.~\eqref{eq:integrator}) can be rewritten as
\begin{equation}
        \frac{m_i^n}{\Delta t}(\hat{\bm{v}}_i^{n+1}-\bm{v}_i^n)
        =(\bm{f}_i^{int})^{n+1}+(\bm{f}_i^{ext})^{n+1}.
        \label{eq:implicit-integrator}
\end{equation}
Directly solving this nonlinear system may be prone to overshooting issues \citep{gast2015optimization,wang2020hierarchical}. Therefore, we convert it into an equivalent optimization problem and solve it using Newton's method with a line search to promote rapid and robust convergence. If gravity is the only external force, we can construct an energy function, called the incremental potential \citep{kane2000variational}, as
\begin{align}
        E(\hat{\bm{v}}^{n+1})
        &=\frac{1}{2}(\hat{\bm{v}}^{n+1}-\bm{v}^n)^T\bm{M}(\hat{\bm{v}}^{n+1}-\bm{v}^n) + \Phi(\bm{F}^{n+1}(\hat{\bm{v}}^{n+1}))+\Phi^{ext}(\hat{\bm{v}}^{n+1})\\
        &=\frac{1}{2}(\hat{\bm{v}}^{n+1}-\bm{v}^n)^T\bm{M}(\hat{\bm{v}}^{n+1}-\bm{v}^n) + \sum_p V_p^0\Psi\left((\bm{I}+\Delta t\nabla\bm{v}_p^{n+1})\bm{F}_p^n\right)+\Phi^{ext}(\hat{\bm{v}}^{n+1}).
        \label{eq:ip_def}
\end{align}
where $\bm{M}=\operatorname{diag}(m_0^n\bm{I},m_1^n\bm{I},\cdots)$ represents the lumped mass matrix composed of all active nodal masses, $\Phi=\sum_p V_p^0\Psi(\bm{F}_p)$ is the total strain energy, and $\Phi^{ext}$ is the external potential. For gravity, $\Phi^{ext}=-\Delta t\sum_{i\in\mathcal{G}}m_i^n\bm{g}\cdot\hat{\bm{v}}_i^{n+1}$, up to a constant independent of the nodal velocities, so that its derivative gives $-\Delta t\bm{f}_i^{ext}$. From Eq.~\eqref{eq:velocity-gradient}, we obtain the variation
\begin{equation}
        \delta\bm{F}_p^{n+1}
        =\Delta t\left(\delta\hat{\bm{v}}_i^{n+1}(\nabla N_{ip})^T\right)\bm{F}_p^n
        =\Delta t\,\delta\hat{\bm{v}}_i^{n+1}\bm{a}_{ip}^{T},\qquad
        \bm{a}_{ip}=(\bm{F}_p^n)^T\nabla N_{ip},
        \label{eq:deformation_variation}
\end{equation}
for a variation of the velocity at node $i$, with all particle positions and kernel gradients fixed. The normalized shape functions already account for averaging between the two staggered grids. 

Thus, the gradient of the incremental potential w.r.t. the nodal DOFs $\hat{\bm{v}}^{n+1}$ can be computed using the chain rule as
\begin{align}
        \bm{r}_i(\hat{\bm{v}}^{n+1})
        :=\frac{\partial E}{\partial\hat{\bm{v}}_i^{n+1}}
        &=m_i^n(\hat{\bm{v}}_i^{n+1}-\bm{v}_i^n)
        +\Delta t\sum_pV_p^0\bm{P}(\bm{F}_p^{n+1})\bm{a}_{ip}
        -\Delta t(\bm{f}_i^{ext})^{n+1},
        \label{eq:ip_grad}
\end{align}
where $\bm{P}=\partial\Psi/\partial\bm{F}$ is the first Piola--Kirchhoff stress. The discrete internal force is therefore
\begin{equation}
        (\bm{f}_i^{int})^{n+1}
        =-\sum_pV_p^0\bm{P}(\bm{F}_p^{n+1})(\bm{F}_p^n)^T\nabla N_{ip}.
        \label{eq:ck_internal_force}
\end{equation}
Equating Eq.~\eqref{eq:ip_grad} to zero gives the discrete momentum equation in Eq.~\eqref{eq:implicit-integrator}. Thus, minimizing the incremental potential is equivalent to finding a momentum-balanced state of the system. Furthermore, the Hessian matrix of the incremental potential can be computed as
\begin{align}
        (\bm{H}_{ij})_{ab}
        &=m_i^n\delta_{ij}\delta_{ab}
        +\Delta t^2\sum_pV_p^0
        (\mathbb{C}_p)_{aA bB}(\bm{a}_{ip})_A(\bm{a}_{jp})_B,
        \label{eq:ip_hessian}\\
        (\mathbb{C}_p)_{aA bB}
        &=\frac{\partial P_{p,aA}}{\partial F_{p,bB}}.\notag
\end{align}
where $i,j$ index nodes in the composite grid, $a,b$ denote spatial components, and $A,B$ denote material components, with repeated material indices summed. Here $\mathbb{C}_p$ is evaluated at $\bm{F}_p^{n+1}$. 
As an implementation check, we independently verified the analytical residual and Hessian against numerical differentiation, confirming consistency of the implemented first- and second-order derivatives.

\subsection{Hessian Structure and Projected Newton Solver}\label{sec3.4}
The Hessian matrix of the incremental potential in CK-MPM has a special structure. Specifically, if we arrange nodal velocity per grid as $\bm{v}=(\bm{v}_{-}^T,\bm{v}_{+}^T)^T$, where $\bm{v}_{-}=\{\bm{v}_{n_1},\bm{v}_{n_2},\cdots,\bm{v}_{n_s}\}^T$ and $\bm{v}_{+}=\{\bm{v}_{m_1},\bm{v}_{m_2},\cdots,\bm{v}_{m_t}\}^T$ represent the nodal velocities in $\mathcal{G}_{-1}$ and $\mathcal{G}_{+1}$, respectively, the Hessian matrix can be written as
\begin{equation}
        \bm{H}_{ckmpm}=\begin{pmatrix}
                \bm{A}_{s\times s} & \bm{B}_{s\times t}\\
                \bm{B}^T_{s\times t} & \bm{C}_{t\times t}
        \end{pmatrix},
\end{equation}
where $\bm{A}$ and $\bm{C}$ are the Hessian blocks corresponding to grid $\mathcal{G}_{-1}$ and $\mathcal{G}_{+1}$, respectively, and $\bm{B}$ is the off-diagonal coupling between these two grids. These cross-grid blocks arise naturally because a particle's deformation gradient is reconstructed using shape functions from both sets of nodes. For comparison, we use $k$ to represent the size of the Hessian matrix in conventional MPM (i.e., $\bm{H}_{mpm}$ is of size $k\times k$). With sufficiently large resolution and the same grid spacing $\Delta x$ for the two methods, the width of kernel function will not change the number of active nodes significantly. This implies that $s\approx t\approx k$, i.e., the size of $\bm{H}_{ckmpm}$ is approximately twice that of $\bm{H}_{mpm}$.

\begin{figure}[htbp]
    \centering
    \begin{subfigure}[c]{0.25\linewidth}
        \centering
        \includegraphics[width=\linewidth]{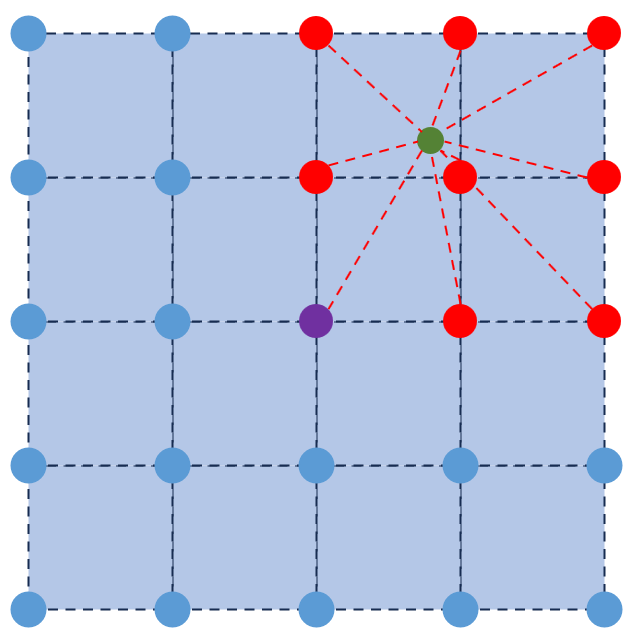}
        \caption{Quadratic Kernel}
    \end{subfigure}
    \hspace{0.1\linewidth}
    \begin{subfigure}[c]{0.5\linewidth}
        \centering
        \includegraphics[width=\linewidth]{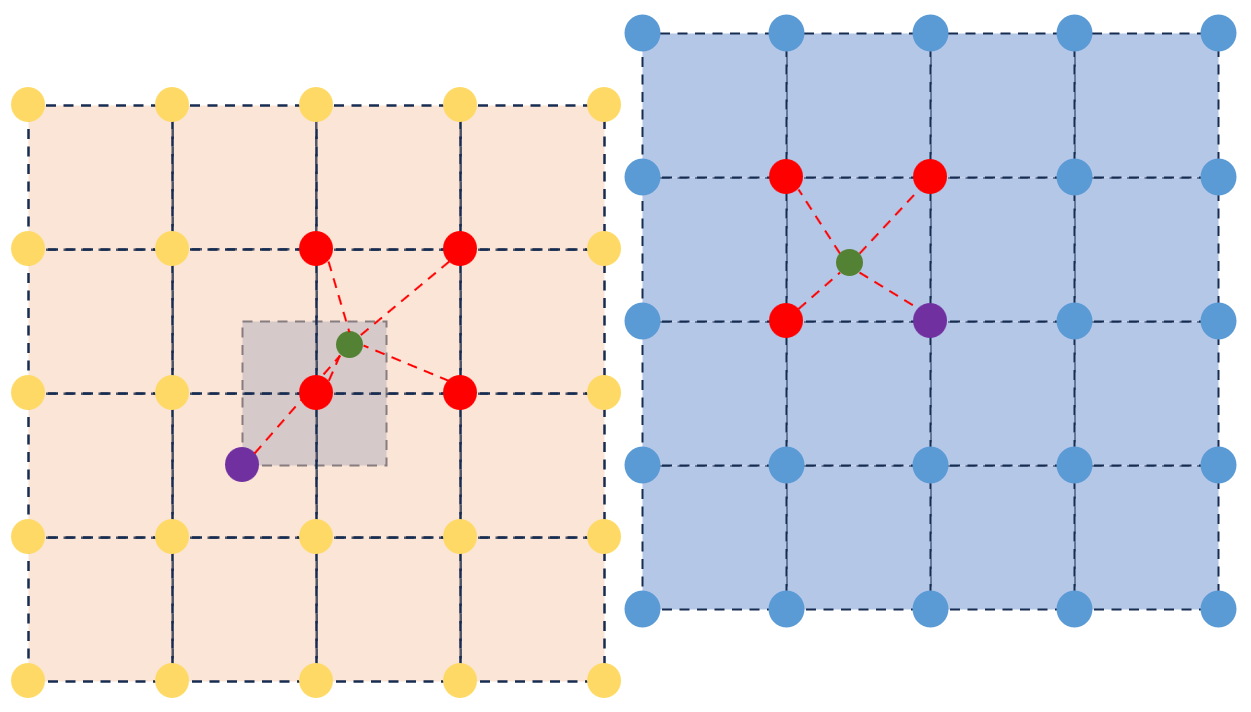}
        \caption{Compact Kernel}
    \end{subfigure}
    \caption{Node coupling induced by particles for quadratic B-spline MPM and CK-MPM. Two grid nodes contribute to the same Hessian block row if they share at least one particle. In 2D, a node in quadratic B-spline MPM is coupled to 24 other nodes, giving 25 nonzero blocks per Hessian row when the diagonal block is included. In CK-MPM, a node is coupled to 8 other nodes on the same grid and 16 nodes on the staggered grid, giving $9+16=25$ nonzero blocks per row. In 3D, the corresponding counts are 125 blocks per row for quadratic B-spline MPM and $27+64=91$ blocks per row for CK-MPM.}
    \label{fig:kernel-Hessian}
\end{figure}

The sparsity pattern of the Hessian is directly determined by the nodal coupling induced by shared particles. In a $d\times d$ blockwise representation, the number of nonzero blocks in row $i$ equals the number of nodes, including node $i$ itself, that share at least one particle with node $i$. As shown in \autoref{fig:kernel-Hessian}, for quadratic B-spline MPM, a node is coupled to 24 other nodes in 2D and 124 other nodes in 3D. Including the diagonal block associated with the node itself, this results in 25 and 125 nonzero blocks per row, respectively. For CK-MPM, a node is coupled to 8 other nodes on the same grid and 16 nodes on the staggered grid in 2D, resulting in $9+16=25$ nonzero blocks per row. In 3D, the corresponding numbers are 26 same-grid neighbors and 64 cross-grid nodes, giving $27+64=91$ nonzero blocks per row. Therefore, although $\bm{H}_{ckmpm}$ has a larger dimension than $\bm{H}_{mpm}$, it contains fewer nonzero blocks per row in 3D. This per-row sparsity does not by itself imply a smaller total Hessian or lower linear-solver cost, since the composite grid has approximately twice as many nodes. Together with the compact particle--grid transfer stencil, this sparsity provides useful structure for efficient implementation of implicit CK-MPM. A systematic comparison and further optimization of its overall computational efficiency are left for future work.

We solve the nonlinear system at each implicit time step using a projected Newton method. At each Newton iteration, the local elastic Hessian contributions are projected onto the positive-semidefinite cone before assembly, yielding a positive-semidefinite global tangent matrix when each complete particle-level contribution, including its cross-grid coupling, is projected consistently. Dirichlet boundary conditions are enforced independently on the active nodes of both staggered grids by projecting the corresponding residual and Hessian degrees of freedom, thereby fixing the prescribed components in the Newton solve. For the dynamic solver, the Hessian matrix remains nonsingular due to the inertial mass contribution on the active, unconstrained nodal degrees of freedom. The resulting linear system is solved using preconditioned conjugate gradients with a vertex-wise $d\times d$ block-Jacobi preconditioner (a $3\times3$ block in 3D), where each diagonal block is inverted independently. PCG uses a relative residual tolerance of $10^{-6}$ and a maximum of $30{,}000$ iterations. We employ an Armijo backtracking line search for global convergence and terminate the Newton iteration once the relative infinity norm of the gradient falls below $10^{-4}$. The line search starts from a unit step and reduces the step by a factor of $0.9$ until both deformation feasibility and the Armijo condition are satisfied. If the step falls below $10^{-3}$, the Newton iteration terminates and retains the last accepted iterate. An infeasible initial Newton iterate is reset to zero velocity before the solve. The whole optimization method we used to solve the implicit time integration problem is outlined in Algorithm~\autoref{alg:projected_newton}.

\begin{algorithm}[htbp]
\caption{Projected Newton Method for Implicit Time Integration}
\label{alg:projected_newton}
\begin{algorithmic}[1]
\Require Grid velocity $\bm{v}^n$, time step $\Delta t$
\Ensure Updated grid velocity $\hat{\bm{v}}^{n+1}$

\State \textbf{Parameters:} $\beta = 0.9$, $c = 10^{-4}$,
$\varepsilon = 10^{-4}$, $\alpha_{\min}=10^{-3}$
\State Initialize $\bm{v}^{(0)} \gets \bm{v}^n$
\If{$\bm{v}^{(0)}$ is infeasible}
    \State Reset $\bm{v}^{(0)}$ to zero velocity
\EndIf
\State Evaluate $\bm{r}^{(0)} = \nabla E(\bm{v}^{(0)})$

\For{$k = 0,1,2,\dots$ until convergence}

    \State Evaluate residual
    \[
        \bm{r}^{(k)} = \nabla E(\bm{v}^{(k)})
    \]

    \If{$
    \|\bm{r}^{(k)}\|_{\infty}
    /
    \|\bm{r}^{(0)}\|_{\infty}
    < \varepsilon
    $}
        \State \textbf{break}
    \EndIf

    \State Evaluate local Hessian blocks and project them to PSD
    \State Assemble the projected global Hessian $\widetilde{\bm{H}}^{(k)}$
    \State Apply Dirichlet boundary conditions

    \State Solve with PCG (relative tolerance $10^{-6}$, cap $30{,}000$):
    \[
        \widetilde{\bm{H}}^{(k)}
        \Delta \bm{v}^{(k)}
        =
        -\bm{r}^{(k)}
    \]

    \State Initialize step size $\alpha \gets 1$

    \While{the trial deformation is infeasible or
    $E(\bm{v}^{(k)} + \alpha \Delta \bm{v}^{(k)})
    >
    E(\bm{v}^{(k)})
    +
    c\,\alpha\,
    \bm{r}^{(k)T}\Delta \bm{v}^{(k)}
    $}
        \State $\alpha \gets \beta \alpha$
        \If{$\alpha < \alpha_{\min}$}
            \State \textbf{break}
        \EndIf
    \EndWhile
    \If{$\alpha < \alpha_{\min}$}
        \State \textbf{break} \Comment{Retain the last accepted iterate}
    \EndIf

    \State Update
    \[
        \bm{v}^{(k+1)}
        \gets
        \bm{v}^{(k)}
        +
        \alpha \Delta \bm{v}^{(k)}
    \]

\EndFor

\State $\hat{\bm{v}}^{n+1} \gets \bm{v}^{(k)}$

\end{algorithmic}
\end{algorithm}

\subsection{Statics Solver}\label{sec3.5}
The implicit formulation can be extended naturally to solve directly for static equilibrium in a single numerical optimization. Specifically, we ignore the inertia term in the incremental potential, together with $\Delta t$ and $\bm{v}_i$, and instead minimize the total potential energy w.r.t. $\Delta\bm{x}$:
\begin{align}
        E(\Delta\bm{x})
        &=\Phi(\bm{x}+\Delta\bm{x})+\Phi^{ext}(\Delta\bm{x})\\
        &=\sum_pV_p^0\Psi\left((\bm{I}+\nabla(\Delta\bm{x}_p))\bm{F}_p\right)
        +\Phi^{ext}(\Delta\bm{x}).
        \label{eq:objective_statics}
\end{align}
Similar to time integration, we have
\begin{equation}
        \nabla(\Delta\bm{x}_p)
        =\sum_{i\in\mathcal{G}}\Delta\bm{x}_i(\nabla N_{ip})^T,
\end{equation}
\begin{equation}
        \delta\bm{F}_p'
        =\delta\Delta\bm{x}_i\bm{a}_{ip}^T,
        \qquad\bm{a}_{ip}=(\bm{F}_p)^T\nabla N_{ip},
\end{equation}
where $\bm{F}_p'=(\bm{I}+\nabla(\Delta\bm{x}_p))\bm{F}_p$. Then, using the chain rule, the gradient of the objective function (Eq.~\eqref{eq:objective_statics}) can be computed as
\begin{equation}
        \frac{\partial E}{\partial\Delta\bm{x}_i}
        =\sum_pV_p^0\bm{P}(\bm{F}_p')\bm{a}_{ip}-\bm{f}_i^{ext},
\end{equation}
where the external potential is chosen consistently so that $\partial\Phi^{ext}/\partial\Delta\bm{x}_i=-\bm{f}_i^{ext}$. Similarly, the stationary point of $\nabla E$ corresponds to the solution of the static equilibrium equation $\bm{f}^{int}+\bm{f}^{ext}=\bm{0}$, with $\bm{f}^{int}$ defined as the negative derivative of the internal strain energy. The Hessian matrix of this objective function can be computed as
\begin{equation}
        (\bm{H}_{ij}^{static})_{ab}
        =\sum_pV_p^0(\mathbb{C}_p')_{aA bB}
        (\bm{a}_{ip})_A(\bm{a}_{jp})_B,
        \qquad
        (\mathbb{C}_p')_{aA bB}
        =\frac{\partial P_{p,aA}(\bm{F}_p')}{\partial F_{p,bB}'},
\end{equation}
which has the same elastic structure as the dynamic Hessian but without the inertia term and its $\Delta t^2$ scaling. One caveat is that the Hessian matrix here is more likely to be indefinite or singular as it does not include the mass matrix. In the static solve, besides projecting each local Hessian to PSD, we therefore add a small diagonal regularization $\epsilon\bm{I}$ to improve numerical conditioning and robustness, together with the prescribed displacement boundary conditions. After solving for $\Delta\bm{x}$ on the grid, the position and deformation gradient of the particles can be updated by
\begin{equation}
        \bm{x}_p^s=\bm{x}_p^0+\sum_{i\in\mathcal{G}}\Delta\bm{x}_i N_{ip},
\end{equation}
and
\begin{equation}
        \bm{F}_p^s
        =(\bm{I}+\nabla(\Delta\bm{x}_p))\bm{F}_p^0
        =\left(\bm{I}+\sum_{i\in\mathcal{G}}\Delta\bm{x}_i(\nabla N_{ip})^T\right)\bm{F}_p^0.
\end{equation}
Here, superscripts $s$ and $0$ denote the static and initial states of the system, respectively. As in the dynamic formulation, the normalized compact-kernel shape functions allow the static equilibrium problem to be written in the standard MPM variational form.

\section{Numerical Verification}

\subsection{2D Cantilever Beam}

To verify the implementation of CK-MPM, we consider the two-dimensional self-weight bending of a cantilever beam. This benchmark is widely used for assessing numerical methods for highly deformable soft materials, since the beam undergoes large deflection relative to its thickness. As such, it provides a suitable test for examining the robustness of CK-MPM under large deformation. The numerical predictions are compared against those obtained with conventional MPM based on quadratic B-spline kernels, as well as against the analytical asymptotic bounds reported by \citet{bickley1934heavy} and \citet{shield1992bending}. Since this is a static nonlinear deformation problem, only positions need to be transferred, while momentum transfer is not required.

\begin{figure}[htbp]
        \centering
        \includegraphics[width=0.5\textwidth]{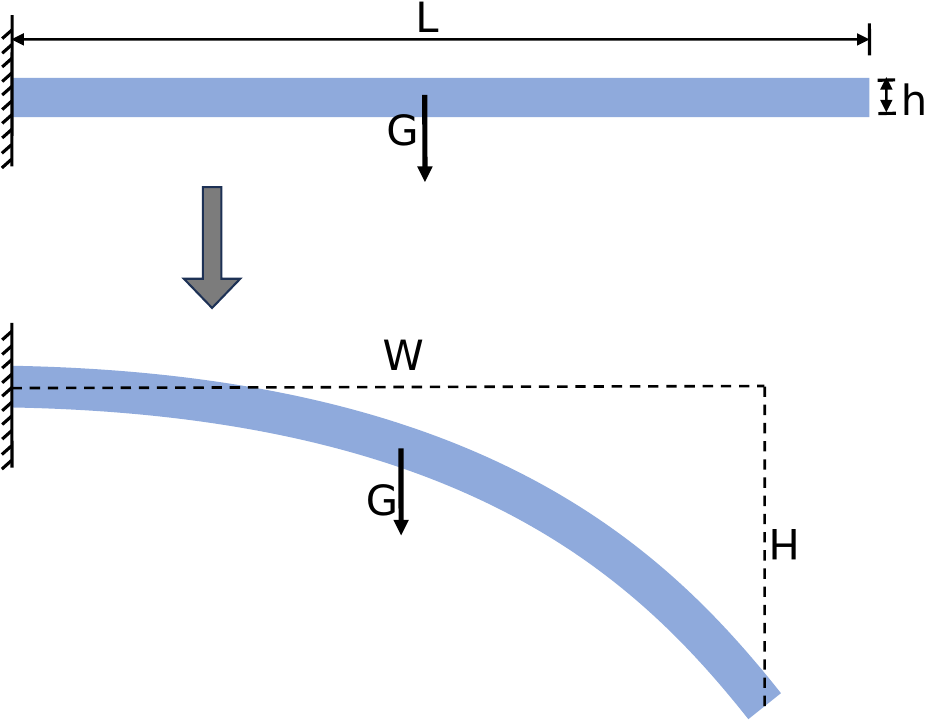}
        \caption{2D cantilever beam configuration. The beam is clamped at the left end and undergoes bending under its self-weight. Here, $L$ and $h$ denote the beam length and thickness, respectively, while $W$ and $H$ represent the horizontal and vertical displacements of the free end.}
        \label{fig:cantilever-setup}
\end{figure}

The computational setup, as shown in \autoref{fig:cantilever-setup}, follows the standard cantilever-beam configuration, in which the left end is fully fixed and the beam deforms solely under gravitational loading. To account for the large-strain response, a hyperelastic constitutive model is employed, specifically the fixed corotated model \citep{stomakhin2012energetically}, and the problem is solved using a static solution procedure. The background grid spacing is set to $\Delta x = 3.906\times 10^{-3}$ mm, with 4 material points per cell. The beam length and height are prescribed as 250$\Delta x$ and 5.5$\Delta x$, respectively, giving a total of 5500 material points. To induce pronounced deformation under self-weight, the material density is set to $\rho = \SI{1e-3}{g/mm^3}$, the Young’s modulus is varied over the range from 600 Pa to 200 kPa, and the Poisson ratio is taken as $\nu=0.1$. Finally, we apply $g = \SI{10}{m/s^2}$ as the gravitational acceleration.

\begin{figure}[htbp]
        \centering
        \includegraphics[width=0.6\textwidth]{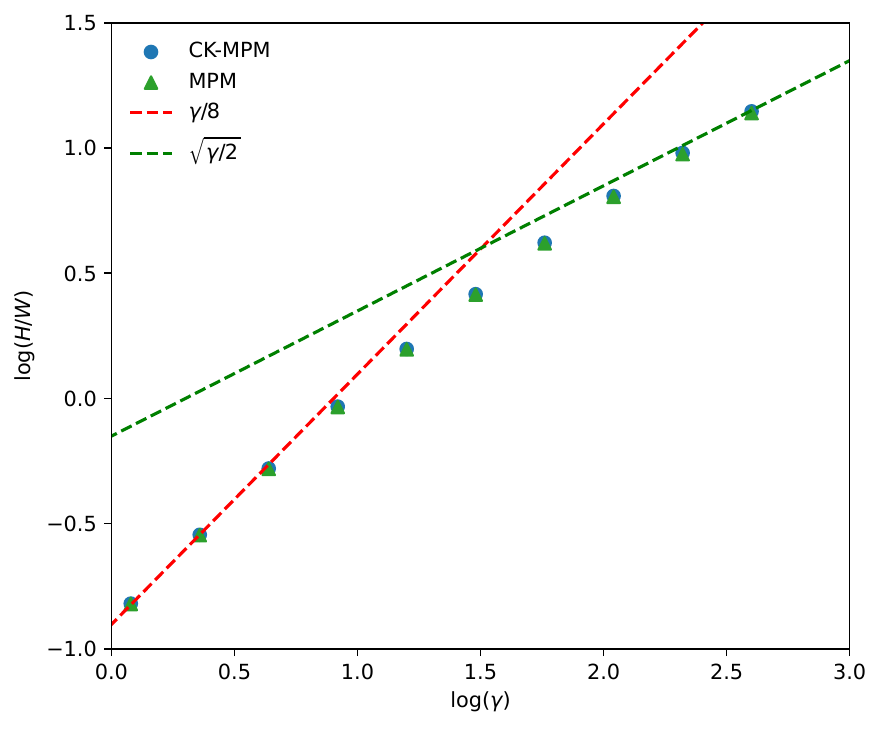}
        \caption{Comparison of the bending ratio $H/W$ versus $\gamma$ obtained with MPM and CK-MPM. The results are nearly indistinguishable, showing that CK-MPM maintains numerical smoothness and predictive accuracy comparable to those of quadratic B-spline MPM. For both methods, the numerical results lie within the analytical asymptotic bounds.}
        \label{fig:cantilever-result}
\end{figure}

Previous studies \citep{bickley1934heavy,shield1992bending} have established that the bending ratio $H/W$ of the cantilever beam can be characterized by a dimensionless gravito-bending parameter $\gamma$ as
\begin{equation}
        \gamma = \frac{12(1-\nu^2)\rho gL^3}{Eh^2},
\label{gamma-hw}
\end{equation}
measuring the relative dominance of gravitational loading over material stiffness. As this parameter increases, corresponding to progressively softer material behavior, the beam response transitions from small deformation towards pronounced large deformation. In the double-logarithmic plane, the solution trend is bounded by two analytical asymptotic lines: one associated with the small-deformation limit $\gamma/8$ and the other defining the upper bound in the large-deformation range $\sqrt{\gamma/2}$. To evaluate the numerical behavior of the proposed method, we plot the predictions of both CK-MPM and quadratic B-spline MPM in the double-logarithmic space $\log({H}/{W})-\log\gamma$. As shown in \autoref{fig:cantilever-result}, the numerical results produced by both methods remain well confined between the analytical asymptotes. Furthermore, the difference between the two methods is negligible, indicating that CK-MPM retains a level of numerical smoothness and predictive accuracy comparable to that of quadratic B-spline MPM, despite using a more compact kernel support.

To further assess the spatial convergence of CK-MPM, we repeat the cantilever simulation at a fixed Young's modulus of $E=50$ kPa while progressively refining the spatial discretization. The same static solution procedure and fixed corotated hyperelastic constitutive model used above are adopted throughout this study. We define the reference grid spacing as $\Delta x^0=1.5626\times10^{-2}$ mm and consider refinement levels $r\in\{1,2,4,8,16\}$, for which the nominal grid spacing is $\Delta x=\Delta x^0/r$. Material points are initialized with a spacing of $0.5\Delta x$. Accordingly, the number of material points along the beam length increases from $125$ to $250$, $500$, $1000$, and $2000$, while that through the beam thickness increases from $6$ to $12$, $24$, $48$, and $96$, respectively. Thus, although the discrete resolution is progressively increased, the physical beam dimensions are held fixed at $L=62.5\Delta x^0$ and $h=3\Delta x^0$.

\begin{figure}[htbp]
        \centering
        \includegraphics[width=0.6\textwidth]{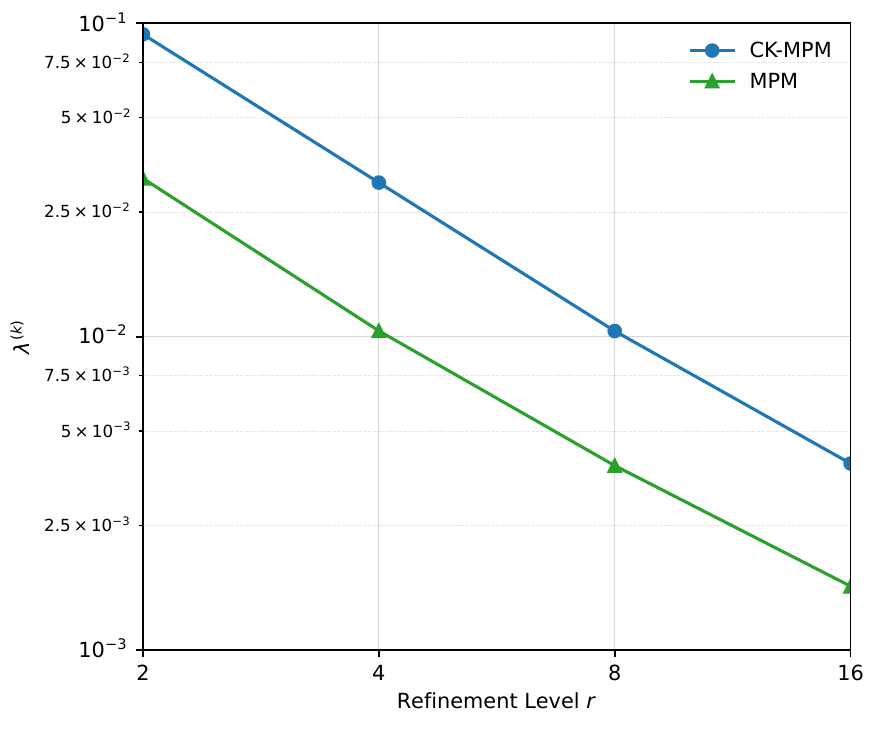}
        \caption{Spatial-refinement convergence of the cantilever beam at $E=50$ kPa. The horizontal axis shows the refinement level $r$, for which $\Delta x=\Delta x^0/r$. The vertical axis shows the relative bending-ratio difference $\lambda^{(k)}$ between two successive refinement levels. CK-MPM converges as the spatial resolution is increased and exhibits the same observed convergence order as quadratic B-spline MPM.}
        \label{fig:cantilever-spatial-refinement}
\end{figure}

For each pair of successive refinement levels $k$ and $k-1$, we define the relative difference in the bending ratio as
\begin{equation}
        \lambda^{(k)} = \frac{\left\lvert(\frac{H}{W})^{(k)}-(\frac{H}{W})^{(k-1)}\right\rvert}{\left\lvert(\frac{H}{W})^{(k-1)}\right\rvert},
        \label{eq:cantilever-spatial-convergence}
\end{equation}
where $(\frac{H}{W})^{(k)}$ is the free-end bending ratio at level $k$, with $H$ and $W$ denoting the vertical and horizontal free-end displacements respectively. As shown in \autoref{fig:cantilever-spatial-refinement}, $\lambda^{(k)}$ decreases systematically as the particle spacing is reduced for both CK-MPM and quadratic B-spline MPM. The two methods exhibit the same observed convergence order, demonstrating that CK-MPM converges under spatial refinement while retaining the convergence behavior of conventional MPM.

\subsection{Hertz Contact Between Cylinder and Rigid Plane }

We next consider a two-dimensional Hertzian contact benchmark consisting of a long cylinder compressed against a rigid plane (\citet{johnson1982one, homel2017field,xiao2021dp}). This example is introduced to assess the contact performance of CK-MPM by comparing the computed stress distribution with both quadratic B-spline MPM and the analytical Hertzian solution. The test is particularly relevant because the contact interface constitutes a strongly discontinuous mechanical interface, at which accurate transmission of contact tractions and faithful resolution of the local deformation field are essential. A linear elastic constitutive model is used in both CK-MPM and quadratic B-spline MPM to ensure a consistent comparison. In this context, a more compact kernel support is expected to reduce artificial contact gaps and mitigate the adverse effects of transfer nonlocality, thereby improving contact accuracy relative to wider-support kernels.

\begin{figure}[htbp]
        \centering
        \includegraphics[width=0.5\textwidth]{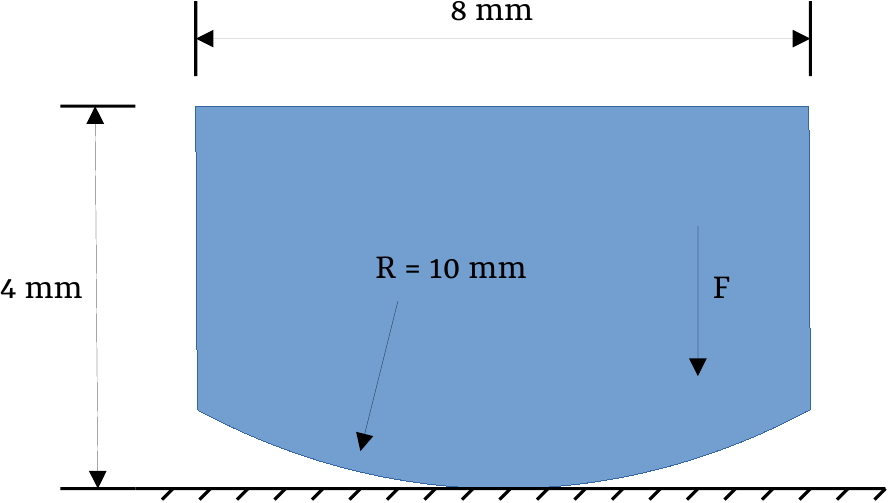}
        \caption{The setup of Hertz contact experiment. Conceptually, it is a cylinder with infinite length put on a rigid plane. We only conduct experiment near the contact area to reduce computational costs.}
        \label{fig:hertz-config}
\end{figure}

The numerical setup is shown in \autoref{fig:hertz-config}. The cylinder radius is set to $R=10$ mm, and the material parameters are adopted from the reference study, with $E=1.0 \times 10^4$ kPa and $\nu=0.07$ (\citet{kakouris2025extended}). The rigid plane is modeled as a frictionless slip boundary: only the normal (vertical) nodal velocity is constrained to zero, while the tangential velocity remains unconstrained. A distributed compressive load of $F=150$ N/m per unit length is applied on the upper surface of the cylinder. As in the previous example, the problem is solved in a quasi-static manner, with a damping coefficient $\zeta = 0.6$ introduced to suppress residual kinetic effects and a time step of $\Delta t=2\times10^{-4}\,\mathrm{s}$. The simulation is initialized in a non-contact configuration, and the load is applied gradually so that the cylinder is slowly pressed onto the rigid plane until a quasi-static state is attained. Both quadratic B-spline MPM and CK-MPM are tested to compare their performance in resolving contact over different contact-patch sizes. To examine accuracy and convergence, four background-grid resolutions are considered, namely $\Delta x = 0.1, 0.05, 0.025,$ and $0.0125$ mm. All cases use $\mathrm{PPC}=9$, corresponding to a particle spacing of $\Delta s=\Delta x/3$ along the cylinder surface. The numerical contact traction is extracted after the quasi-static state is reached and evaluated against the analytical solution. For each particle on the lower outer layer of the cylinder, its normal force $f_y$ is converted to a two-dimensional signed normal traction as $p=f_y/\Delta s=3f_y/\Delta x$. The samples satisfying $|x|\leq a$ are sorted by their horizontal coordinate and compared with the analytical profile. The analytical signed normal-traction distribution is given as 
\begin{equation}
p = -\sigma_{yy}^{\text{max}} \sqrt{1 - (\frac{x}{a})^2}, \space \text{for }\space -a\leq x \leq a ,
\end{equation}
where 
\begin{equation}
{a = 2\sqrt{\frac{FR(1-\nu^2)}{\pi E}},\space \sigma^{\text{max}}_{yy} = \frac{2F}{\pi a}}.
\label{eq:width-a}
\end{equation}

\begin{figure}[htbp]
        \centering
        \includegraphics[width=0.5\textwidth]{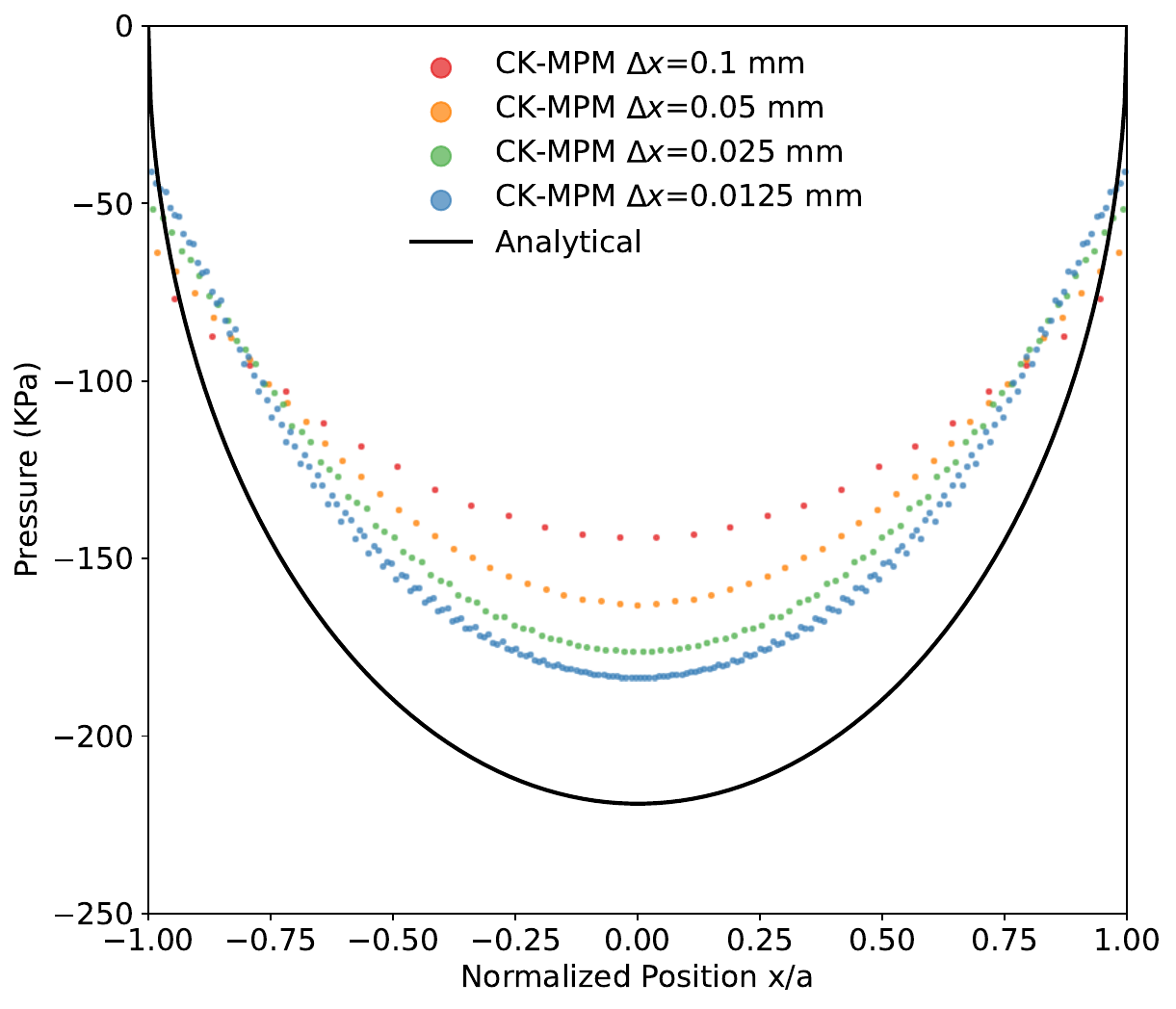}
        \caption{Pressure distribution along the contact interface predicted by CK-MPM for different grid spacings \(\Delta x\), compared with the analytical solution. As \(\Delta x\) decreases, the numerical results progressively converge toward the analytical pressure distribution.}
        \label{fig:hertz-pressure}
\end{figure}

\begin{figure}[htbp]
    \centering

    \begin{subfigure}{0.45\linewidth}
        \centering
        \includegraphics[width=\linewidth]{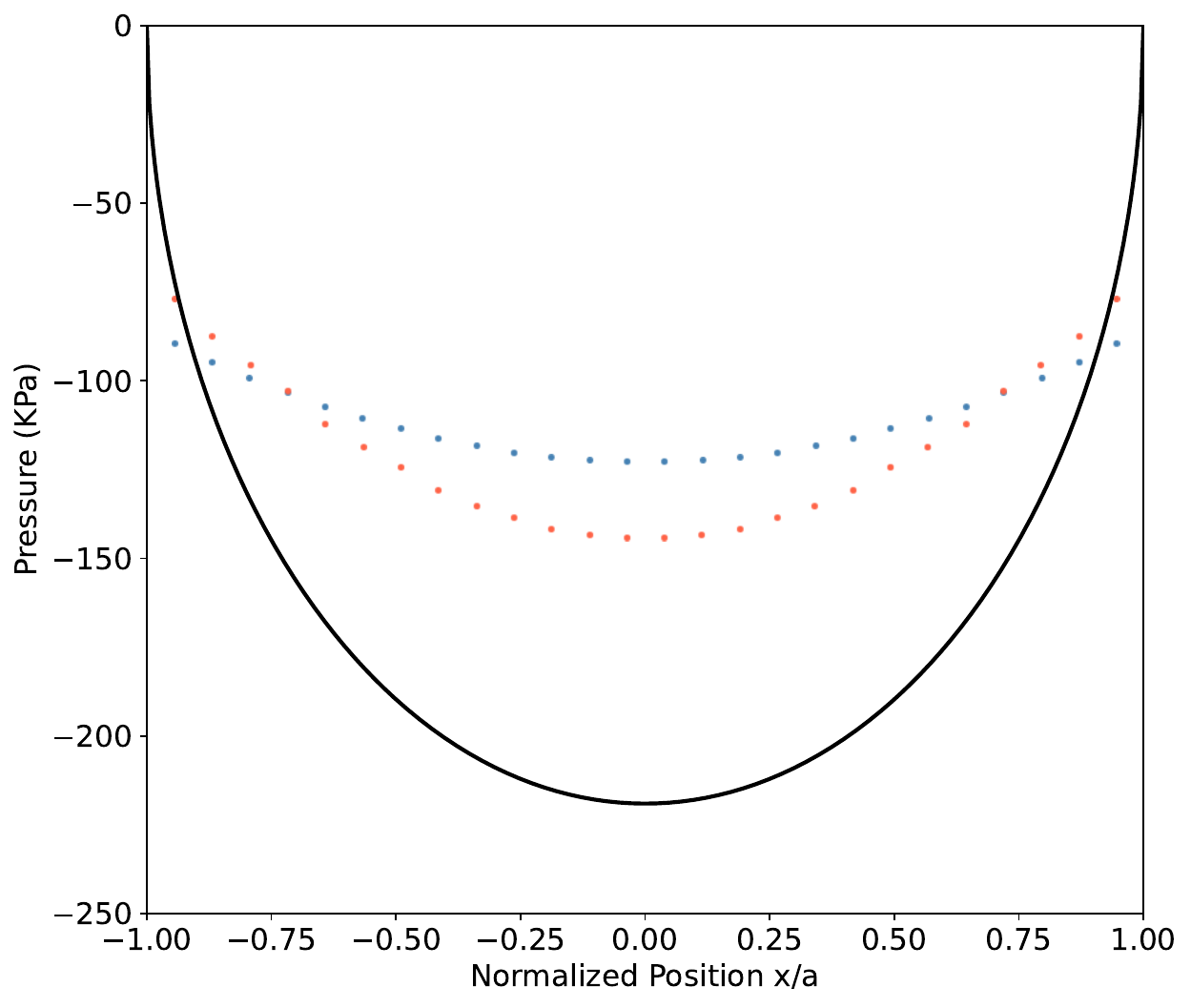}
        \caption{$\Delta x = \SI{0.1}{mm}$}
    \end{subfigure}
    \hfill
    \begin{subfigure}{0.45\linewidth}
        \centering
        \includegraphics[width=\linewidth]{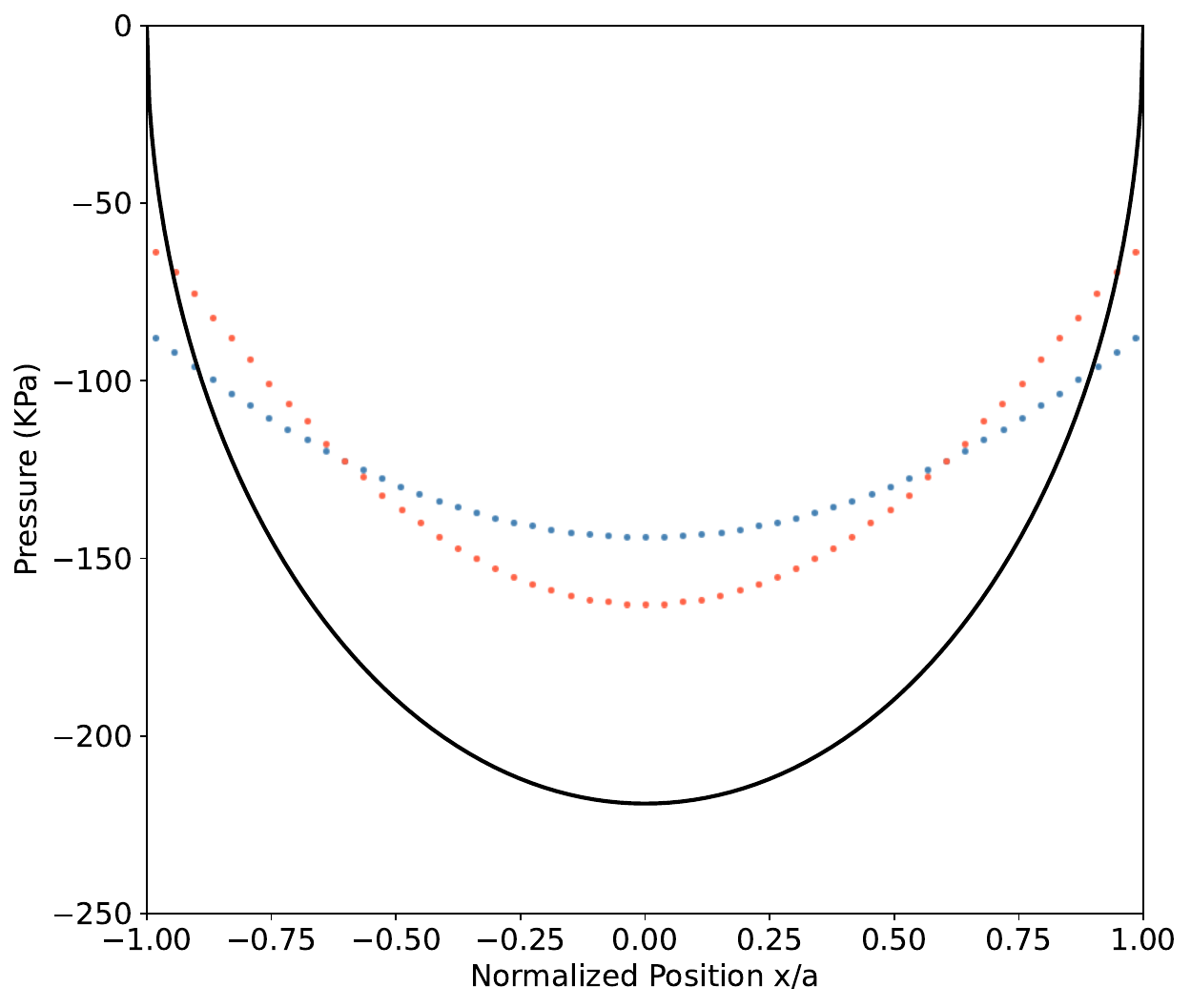}
        \caption{$\Delta x = \SI{0.05}{mm}$}
    \end{subfigure}
    
    \begin{subfigure}{0.45\linewidth}
        \centering
        \includegraphics[width=\linewidth]{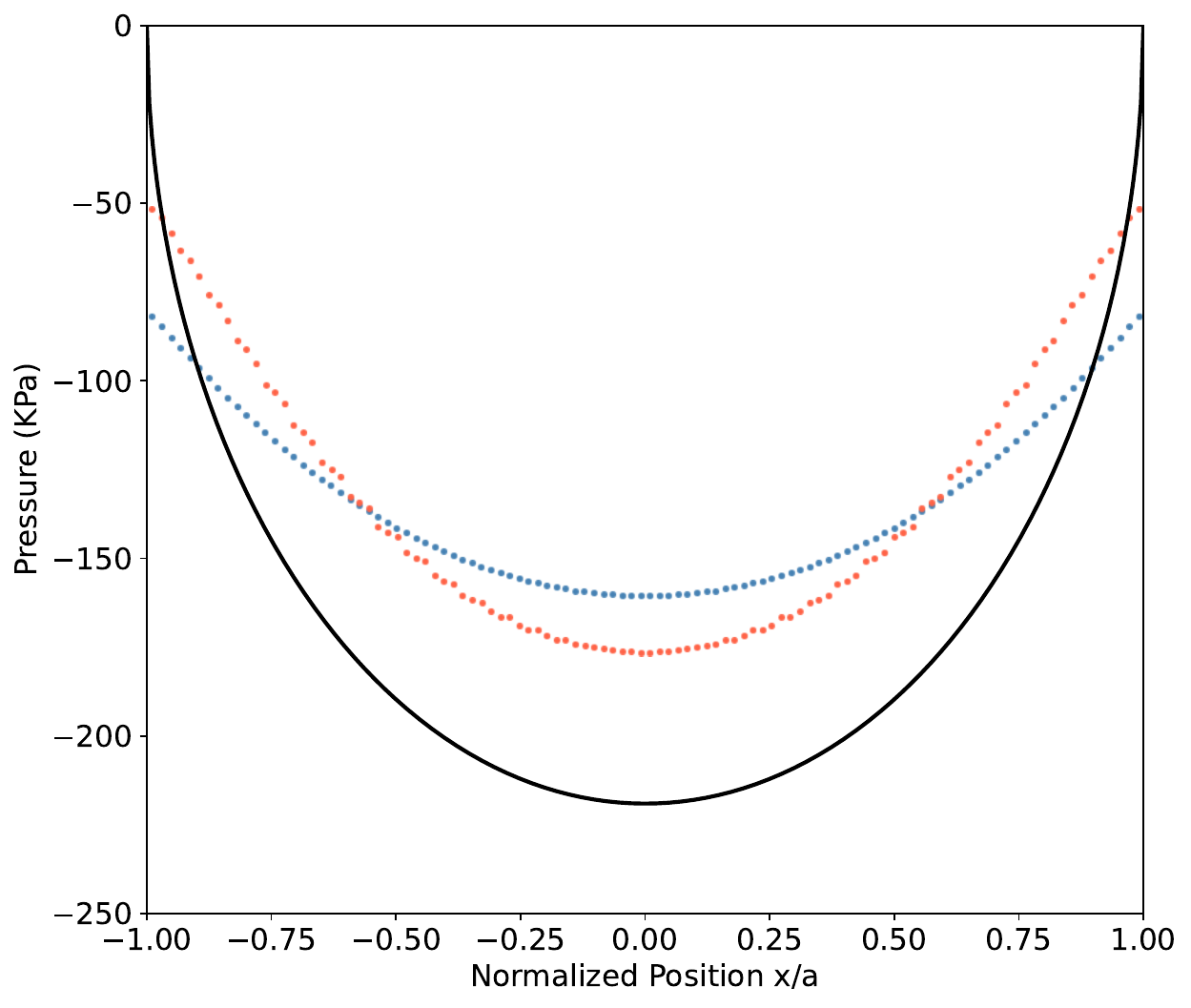}
        \caption{$\Delta x = \SI{0.025}{mm}$}
    \end{subfigure}
    \hfill
    \begin{subfigure}{0.45\linewidth}
        \centering
        \includegraphics[width=\linewidth]{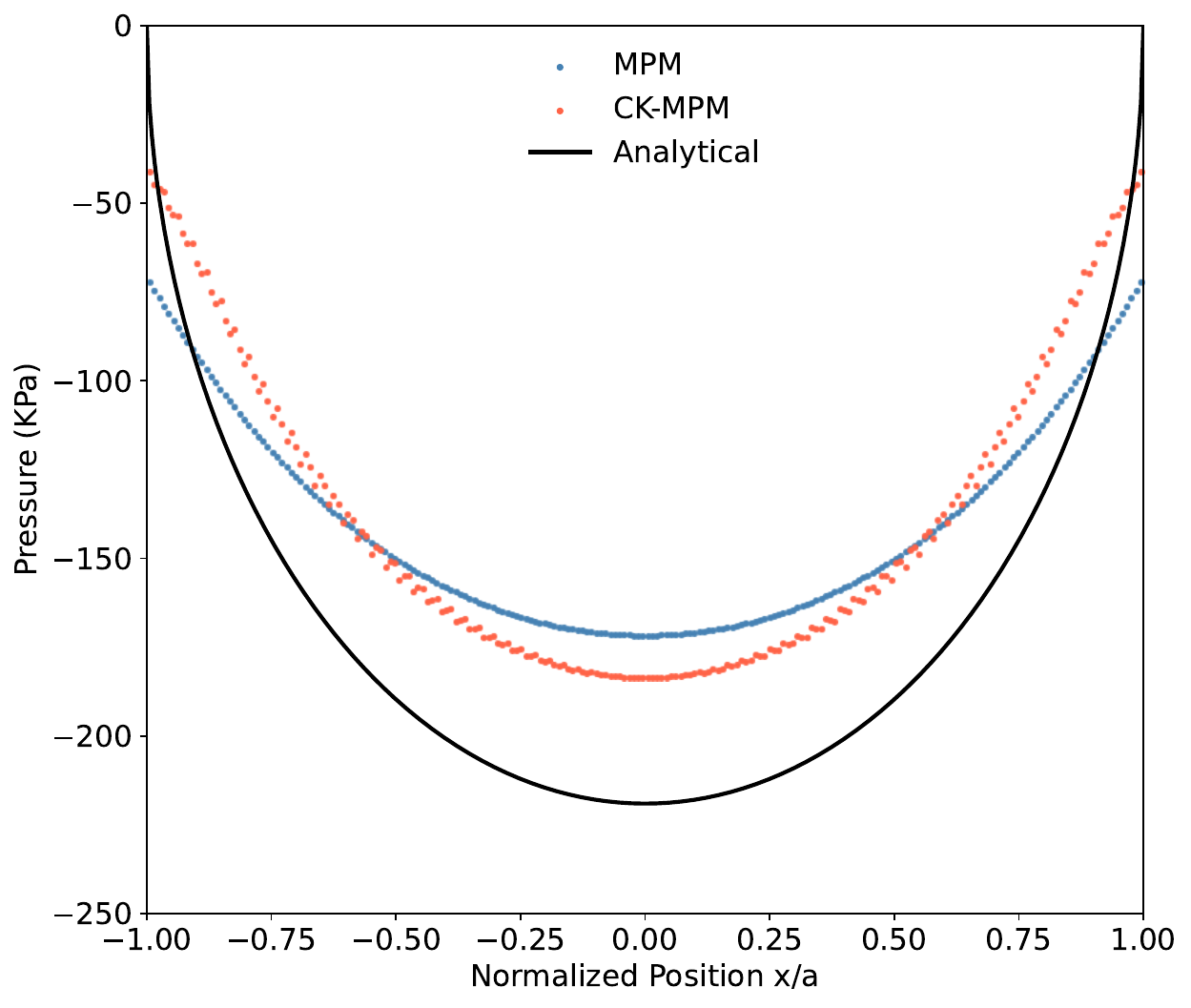}
        \caption{$\Delta x = \SI{0.0125}{mm}$}
    \end{subfigure}

    \caption{Comparison of pressure distributions between CK-MPM and MPM. As shown in the figure, for each $\Delta x$, the distribution obtained with CK-MPM is closer to the theoretical solution. This improvement is attributed to its smaller support radius.}
    \label{fig:pressure-compare}
\end{figure}

The contact-pressure distribution predicted by CK-MPM is presented in \autoref{fig:hertz-pressure}. As the grid spacing $\Delta x$ decreases, the numerical solution is seen to converge progressively toward the analytical Hertzian pressure distribution, with good agreement obtained at the finest resolution considered. A direct comparison between CK-MPM and quadratic B-spline MPM at each resolution is provided in \autoref{fig:pressure-compare}. In all cases, CK-MPM yields a pressure profile that more closely matches the analytical solution, indicating improved contact accuracy relative to the wider-support quadratic B-spline kernel.

\begin{figure}[htbp]
        \centering
        \includegraphics[width=0.6\textwidth]{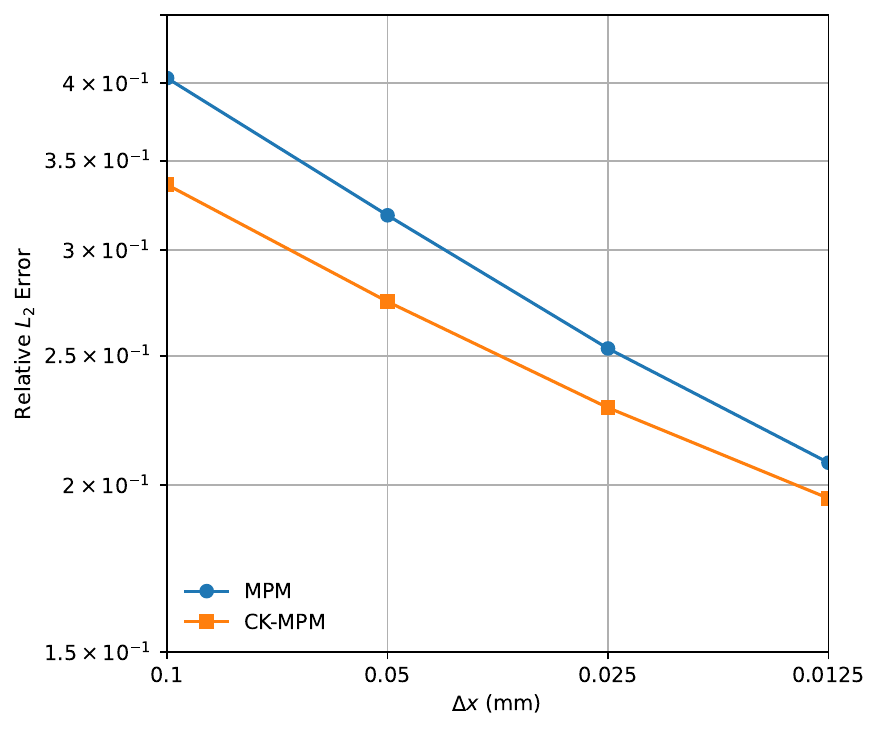}
        \caption{Relative $L_2$ error as a function of \(\Delta x\) for CK-MPM and quadratic MPM. Both methods exhibit a similar convergence trend with refinement, while CK-MPM consistently yields lower errors over the range of \(\Delta x\) considered.}
        \label{fig:hertz-relative-error}
\end{figure}

For a more quantitative assessment, we evaluate the relative $L_2$ error of the pressure distribution over the contact region. The relative $L_2$ error is defined as
\begin{equation}
        e_{L_2} = \left(\frac{\int_{\Gamma_c} \left(p-p_{\mathrm{a}}\right)^2\,\mathrm{d}x}{\int_{\Gamma_c} p_{\mathrm{a}}^2\,\mathrm{d}x}\right)^{1/2},
\end{equation}
where $\Gamma_c$ denotes the contact region, and $p$ and $p_{\mathrm{a}}$ are the numerical and analytical contact pressures, respectively. In the discrete evaluation, the integrals are approximated over the contact samples using the trapezoidal rule. The resulting relative $L_2$ errors for CK-MPM and quadratic B-spline MPM are shown in \autoref{fig:hertz-relative-error}. Although the two methods exhibit comparable convergence rates under grid refinement, CK-MPM consistently achieves lower errors. This improvement is mainly attributed to the more accurate prediction of the contact radius $a$ in Eq.~\eqref{eq:width-a}, which suggests that the compact kernel support reduces the artificial contact gap and thereby enhances the accuracy of contact-pressure transmission.

\begin{figure}[htbp]
        \centering
        \includegraphics[width=\textwidth]{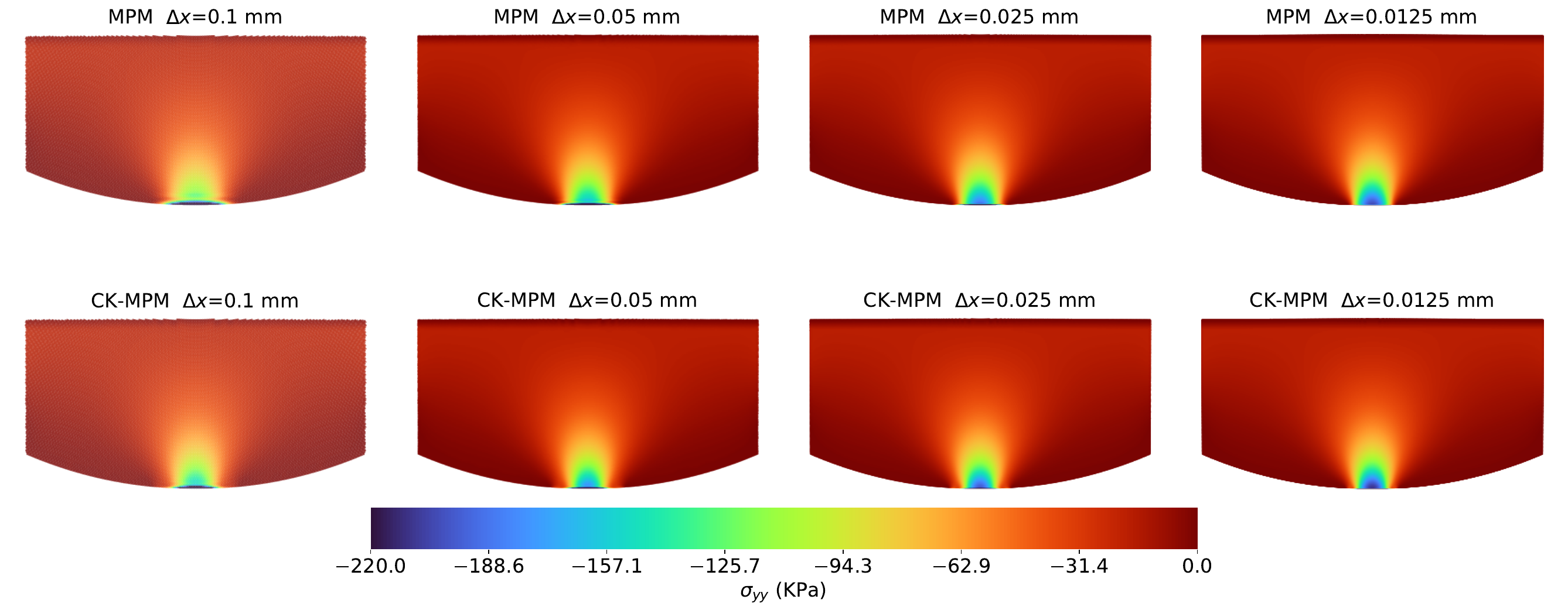}
        \caption{Global stress distributions \(\sigma_{yy}\) obtained with MPM and CK-MPM at different grid spacings \(\Delta x\). The two methods produce comparable global stress fields across all resolutions, indicating that the proposed method preserves the overall stress distribution while providing improved accuracy in the contact region.}
        \label{fig:hertz-stress}
\end{figure}

We further compare the stress fields predicted by the two methods over the entire body, as shown in \autoref{fig:hertz-stress}. Although the differences become visually negligible at fine resolutions, a clear distinction can still be observed at coarser resolutions, where the effect of kernel support is more pronounced. In particular, CK-MPM predicts a narrower contact radius than quadratic B-spline MPM, which is more consistent with the analytical Hertzian solution. This observation further supports the interpretation that the improved performance of CK-MPM originates from its enhanced transfer locality, showing that a more compact kernel can yield higher contact accuracy, especially when the grid resolution is relatively coarse.

It is important to note that, although a non-negligible discrepancy remains between the CK-MPM prediction and the analytical Hertzian solution, the present results are obtained solely from the intrinsic contact-handling mechanism of the underlying MPM framework, without introducing any additional contact-enhancement technique, such as level-set-based interface reconstruction (\citet{bing2019b,liu2020ils,xiao2021dp}), for more accurate treatment of material-point contact. In this sense, CK-MPM is not intended to compete with, or replace, dedicated contact algorithms developed for MPM. Instead, it should be interpreted as a complementary improvement at the transfer-kernel level, which can be naturally combined with existing contact-enhancement strategies to further reduce geometric mismatch and artificial nonconformity at the contact interface.

\section{Numerical Examples}
\subsection{Numerical Diffusion}

Numerical diffusion is a common artifact in the P2G and G2P transfer processes of quadratic B-spline MPM, since kernel-based mapping inevitably involves weighted summation and averaging of nodal quantities, such as momentum, and therefore introduces a degree of numerical nonlocality. Through repeated transfer cycles, this artificial smoothing may accumulate and progressively smear physical fields carried by the material points. As a representative example, heat exchange between two fluids with distinct temperatures and intrinsically low thermal diffusivity may be significantly overpredicted when numerical diffusion is present. From this perspective, a more compact kernel support is desirable because it helps preserve transfer locality and reduce artificial smoothing. Accordingly, CK-MPM is expected to exhibit improved performance in suppressing transfer-induced numerical diffusion owing to its more confined kernel support.

\begin{figure}[htbp]
    \centering
    \begin{subfigure}[t]{0.37\textwidth}
        \centering
        \vspace{0px}
        \includegraphics[width=\textwidth]{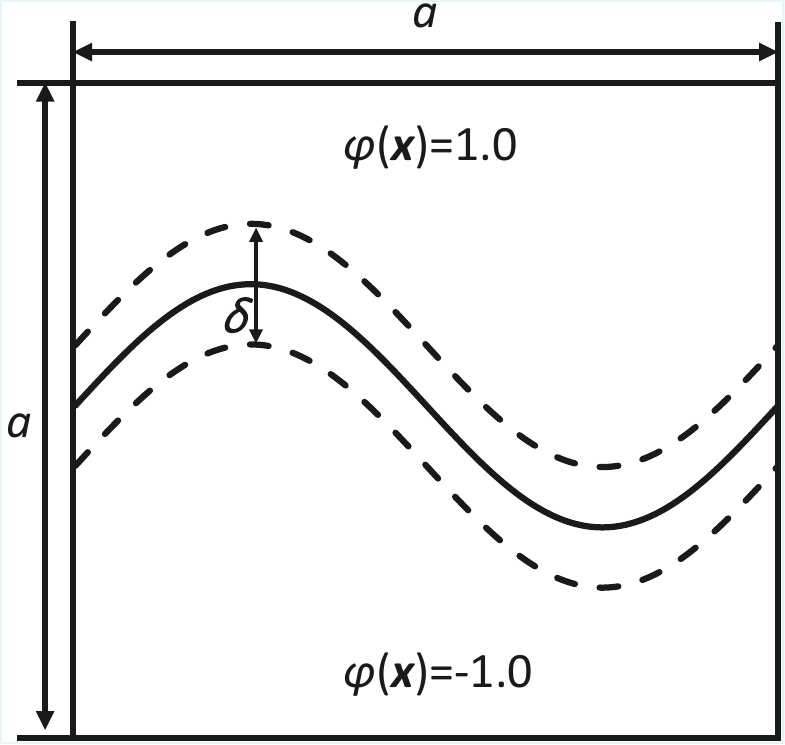}
        \caption{}
    \end{subfigure}
    \hfill
    \begin{subfigure}[t]{0.58\textwidth}
        \centering
        \vspace{0px}
        \includegraphics[width=\textwidth]{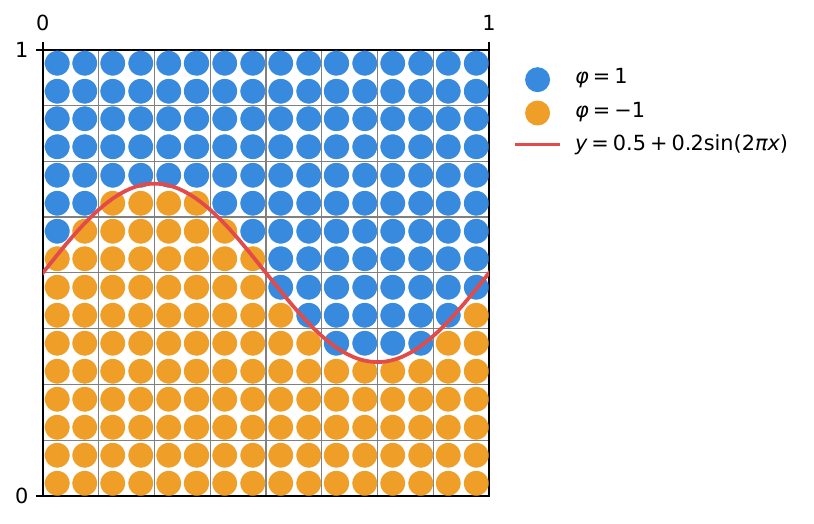}
        \caption{}
    \end{subfigure}

    \caption{Configuration of the numerical diffusion experiment. (a) The values on either side of the interface are different. After transferring iterations a numerical diffusion band with width $\delta$ will appear. In this experiment we use $a = \SI{1}{m}$. (b) The red solid line denotes the interface: particles above it are assigned $\varphi = 1$, while those below are assigned $\varphi = -1$. For visualization purposes, the grid spacing in this figure is $\Delta x = 0.125\,\text{m}$; in the actual experiment, $\Delta x = 0.01\,\text{m}$.}
    \label{fig:diffusion-setup}
\end{figure}

In this section, we present a numerical benchmark to assess the diffusion behavior of CK-MPM in comparison with conventional linear and quadratic B-spline kernels. Since transfer-induced diffusion in MPM originates primarily from the repeated P2G and G2P mappings, the present test is deliberately restricted to the transfer process alone. For all three kernels, the same kernel-weighted P2G/G2P scalar transfer is repeatedly applied, with no governing-equation solve, constitutive model, or time integration. The purpose of this setup is therefore to isolate the sole effect of kernel choice on transfer-induced numerical diffusion. The configuration of the benchmark is shown in \autoref{fig:diffusion-setup}. The computational domain is a square of side length of $1$m, discretized using a uniform background grid. Material points are uniformly distributed over the domain. The sharp interface, defined as $y = 0.5 + 0.2\sin(2\pi x)$, separates the domain into two subregions: particles above the interface are assigned a scalar value of $\varphi(\bm x)=1$, whereas those below are assigned a value of $\varphi(\bm x)=-1$. Numerical diffusion is then generated through repeated P2G and G2P transfer cycles. Its magnitude is quantified by examining the characteristics of the transition band that develops across the initially sharp interface. In practice, we set $\Delta x = \SI{0.01}{m}$ and use $\text{PPC} = 4$. 

\begin{figure}[htbp]
    \centering
    \includegraphics[width=0.9\linewidth]{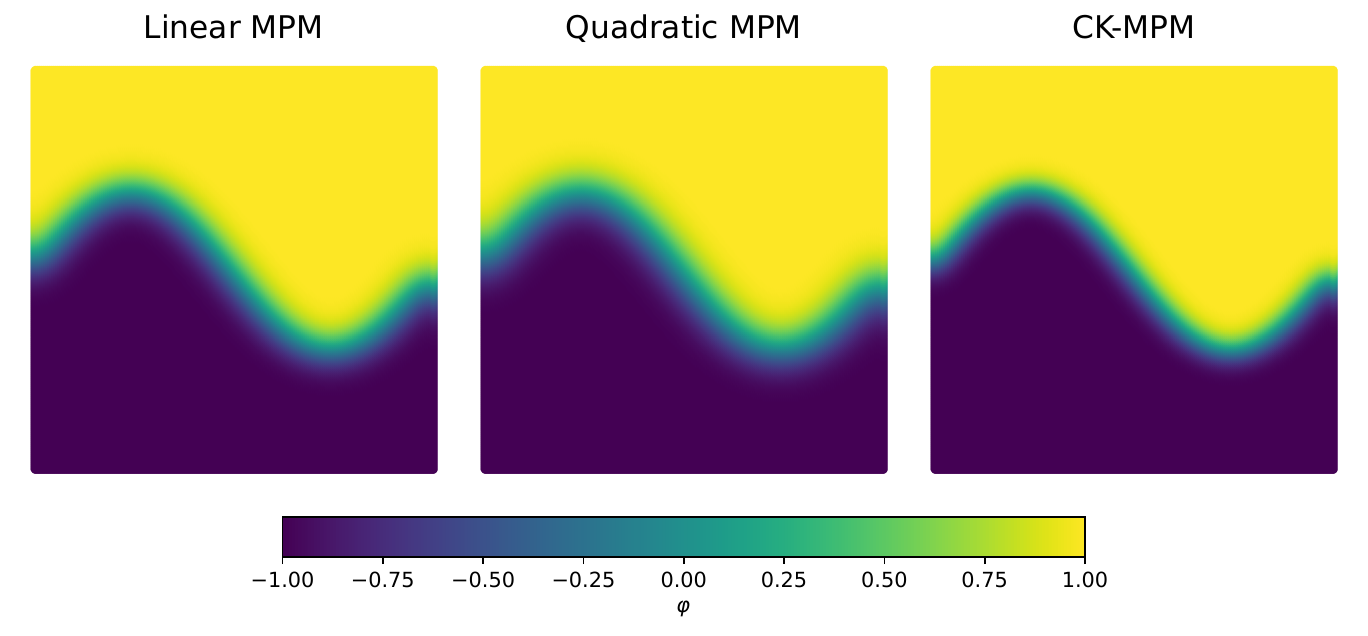}
    \caption{Comparison of numerical diffusion after 50 transfer iterations for different kernels. CK-MPM produces a narrower transition band than the other two kernels.}
    \label{fig:diffusion-compare}
\end{figure}

The distribution of $\varphi$ after 50 transfer iterations is further examined in \autoref{fig:diffusion-compare}. Among the three kernels, CK-MPM produces the narrowest transition band and most effectively preserves the initially sharp separation interface, indicating the weakest transfer-induced numerical diffusion. In comparison, both the linear and quadratic B-spline kernels generate substantially broader transition regions, leading to a more pronounced smearing of the interface.

\begin{figure}[htbp]
    \centering
    \begin{subfigure}[t]{0.45\textwidth}
        \centering
        \includegraphics[width=\textwidth]{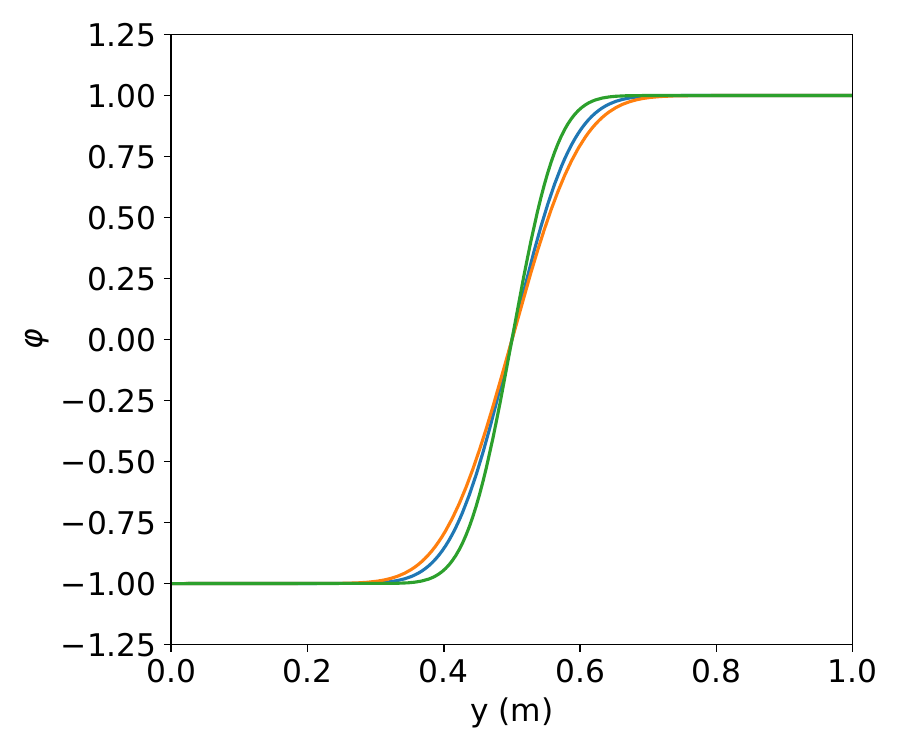}
        \caption{}
        \label{fig:b}
    \end{subfigure}
    \hfill
    \begin{subfigure}[t]{0.53\textwidth}
        \centering
        \includegraphics[width=\textwidth]{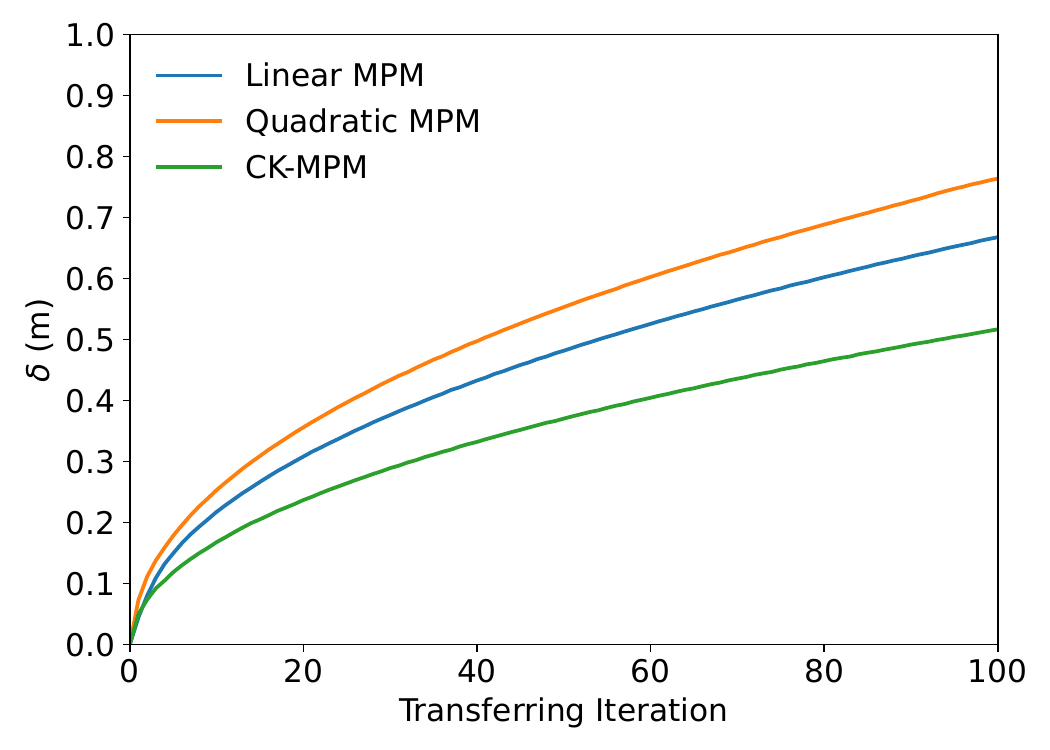}
        \caption{}
    \end{subfigure}

    \caption{(a) One-dimensional slice of particle values along $x = 0.5$ after 50 transfer iterations. Though compact kernel has the same supporting radius with linear kernel, it can reduce diffusion significantly compared with linear kernel. It benefits from more concentration around center area. (b) Numerical diffusion growth in our pure diffusion benchmark. CK-MPM exhibits the slowest growth rate, followed by Linear MPM and Quadratic MPM, demonstrating that compact kernel is able to reduce numerical diffusion phenomenon.}
    \label{fig:diffusion-quant}
\end{figure}

For a more quantitative comparison, we extract the one-dimensional profile along the line $x=0.5$ and measure the evolution of the diffusion width $\delta$, as shown in \autoref{fig:diffusion-quant}(a) and (b), respectively. The results confirm that the adopted compact kernel reduces numerical diffusion in a meaningful way, yielding both the sharpest scalar profile and the lowest growth rate of $\delta$. By contrast, quadratic B-spline MPM shows the poorest performance in preserving interface sharpness. It is also noteworthy that linear MPM, despite having a smaller support size than that of CK-MPM, still exhibits a larger diffusion width and a more smeared interface. This suggests that the improved performance of CK-MPM does not arise solely from support size, but also from the detailed shape of the kernel, which is slightly more localized within the same support radius. This observation is consistent with the kernel design: relative to the quadratic kernel, the compact kernel has a smaller support radius, while relative to the linear kernel, it remains smoother and more centrally concentrated. Overall, these results demonstrate that the compact-kernel design can effectively reduce transfer-induced numerical diffusion within the standard MPM framework.

\subsection{Sphere Falling Through Hollow Cylinder}

We employ the benchmark of a sphere freely falling through a hollow cylinder to demonstrate how transfer-induced numerical nonlocality may fundamentally influence the physical response that the simulation is intended to capture (\citet{liu2025ck}). The example is particularly suitable because the global motion is strongly controlled by local geometric interaction within a narrow clearance. As a result, excessive kernel support may introduce artificial nonlocal smoothing, alter the effective gap, and thereby produce a qualitatively incorrect system response. This benchmark therefore highlights a central advantage of compact kernel support: by preserving transfer locality, it can avoid spurious interactions across narrow geometric gaps and thereby improve the physical fidelity of the predicted dynamics.


\begin{figure}[htbp]
        \centering
        \includegraphics[width=0.3\textwidth]{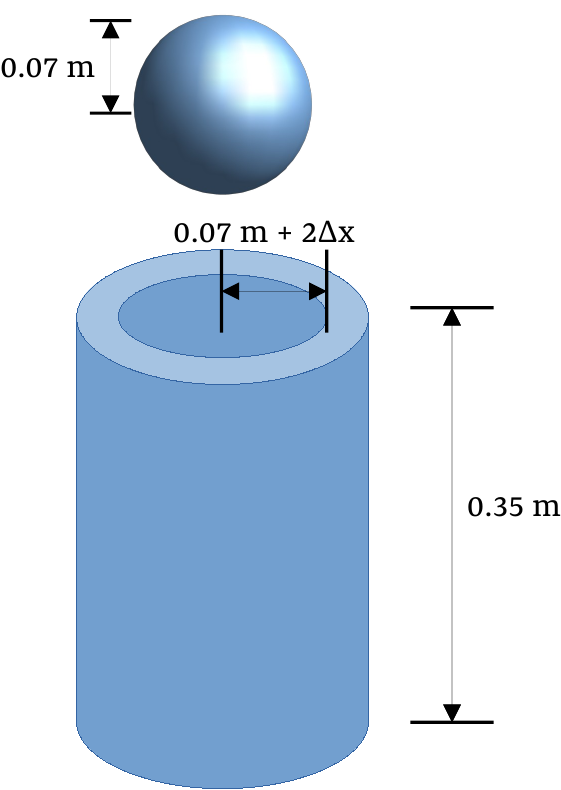}
        \caption{Schematic of the sphere-fall test through a hollow cylinder. The cylinder inner radius is set to be slightly larger than the sphere radius; therefore, the sphere should theoretically pass through the cylinder without contact-induced obstruction. In the numerical tests, CK-MPM reproduces this behavior accurately, whereas quadratic MPM fails to do so.}
        \label{fig:chimney-setup}
\end{figure}

The configuration of this experiment is shown in \autoref{fig:chimney-setup}. A sphere is initially positioned above a hollow cylinder, with its center aligned with the cylinder axis. The sphere has a radius of $0.07\,\mathrm{m}$, while the hollow cylinder has an inner radius of $0.07\,\mathrm{m}+2\Delta x$, providing sufficient clearance for the sphere to fall freely through the cylinder. The two objects are assigned the same material properties, with $E=1.0\times10^5\,\mathrm{kPa}$ and $\nu=0.1$. Since both objects are relatively stiff and undergo only small deformations compared with the characteristic geometric scale, a linear elastic constitutive model is adopted. We compare CK-MPM against quadratic B-spline MPM under identical simulation settings, with both methods employing the APIC transfer scheme so that the comparison primarily reflects the influence of the underlying kernel support. The grid spacing is set to $\Delta x=7.8125\times10^{-3}\,\mathrm{m}$ with PPC $=8$, providing a sufficiently fine geometric representation of the curved surfaces of both the sphere and the inner cylinder. The simulation is initialized by releasing the sphere above the hollow cylinder and allowing it to fall freely under gravity. The time step is $\Delta t=10^{-3}\,\mathrm{s}$.

\begin{figure}[htbp]
        \centering
        \includegraphics[width=0.6\textwidth]{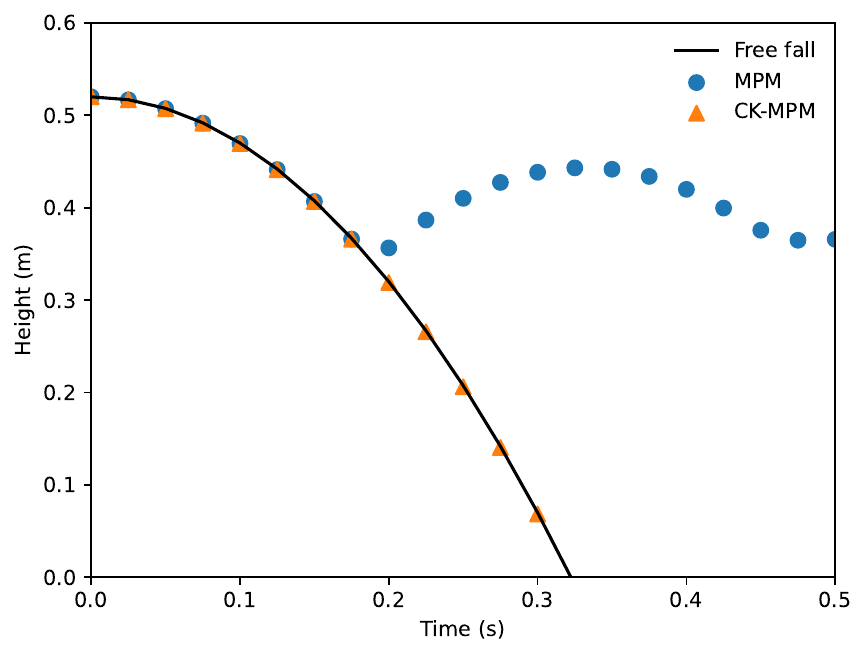}
        \caption{Time history of the sphere center height. Owing to its larger kernel support, quadratic MPM does not correctly reproduce the passage of the sphere through the hollow cylinder. By contrast, the CK-MPM results closely follow the free-fall trajectory, indicating that the sphere motion remains essentially unaffected by spurious contact forces throughout the process.}
        \label{fig:chimney-time}
\end{figure}

\autoref{fig:chimney-time} presents the time history of the vertical position of the sphere center. For the simulation performed with CK-MPM, the sphere passes successfully through the hollow cylinder and its trajectory remains in good agreement with the analytical free-fall solution. By contrast, the quadratic B-spline MPM prediction deviates markedly from the analytical response, indicating that the sphere is unable to pass through the narrow clearance. This discrepancy is primarily attributed to the wider kernel support, which induces artificial numerical contact with the inner surface of the cylinder and consequently generates spurious contact forces that impede the motion of the sphere.

\begin{figure}[htbp]
        \centering
        \includegraphics[width=0.95\textwidth]{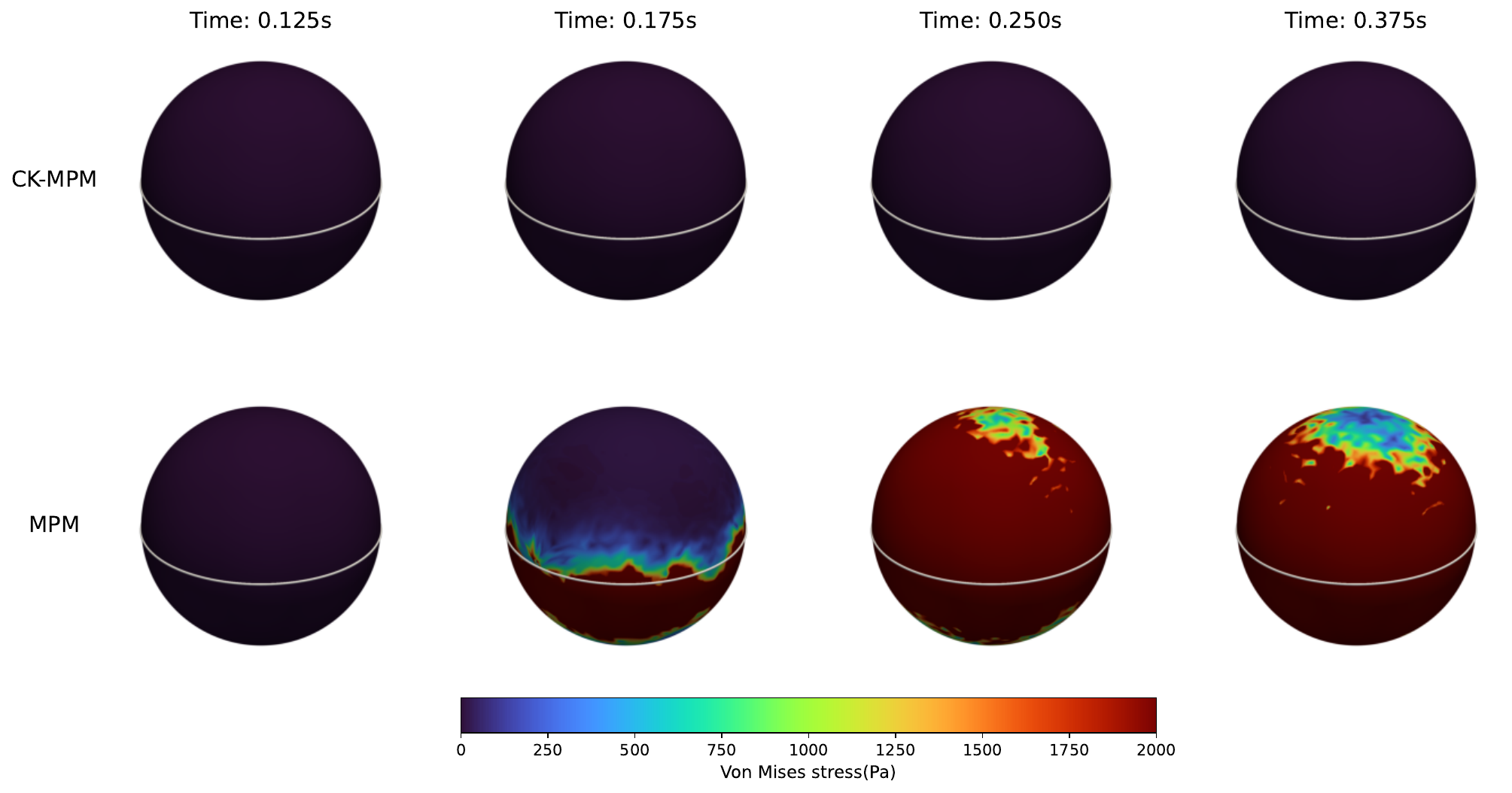}
        \caption{Comparison of the von Mises stress distributions. The sphere computed with MPM exhibits substantially higher stress levels than that computed with CK-MPM, reflecting the stronger spurious contact forces generated between the sphere and the hollow cylinder in MPM.}
    \label{fig:chimney-stress}
\end{figure}

Further evidence is provided in \autoref{fig:chimney-stress}, which shows the von Mises stress distribution on the sphere surface. In the quadratic B-spline MPM result, a clear spurious contact response can be identified between the sphere and the inner wall of the hollow cylinder, despite the existence of sufficient geometric clearance. These results demonstrate that CK-MPM is able to reduce the artificial contact gap through its more compact kernel support, thereby improving physical fidelity in narrow-clearance contact scenarios. This feature is particularly important for simulations in which the correct system response is highly sensitive to local geometric interaction.

\subsection{Impact of Two Neo-Hookean Rings}

We further consider the collision of two equal-sized two-dimensional hyperelastic rings as a dynamic benchmark for evaluating the conservation properties of CK-MPM. This example is particularly demanding because the impact induces large deformation together with rapid and repeated conversion between kinetic energy and elastic strain energy (\citet{liang2024mortar, kakouris2025extended}). As a result, even relatively small numerical errors in particle-grid transfer or time integration may accumulate and become visible in the global energy history. The benchmark therefore provides a stringent test of the ability of CK-MPM to maintain stable and physically consistent energy evolution in highly deformable transient dynamics.

\begin{figure}[htbp]
        \centering
        \includegraphics[width=0.6\textwidth]{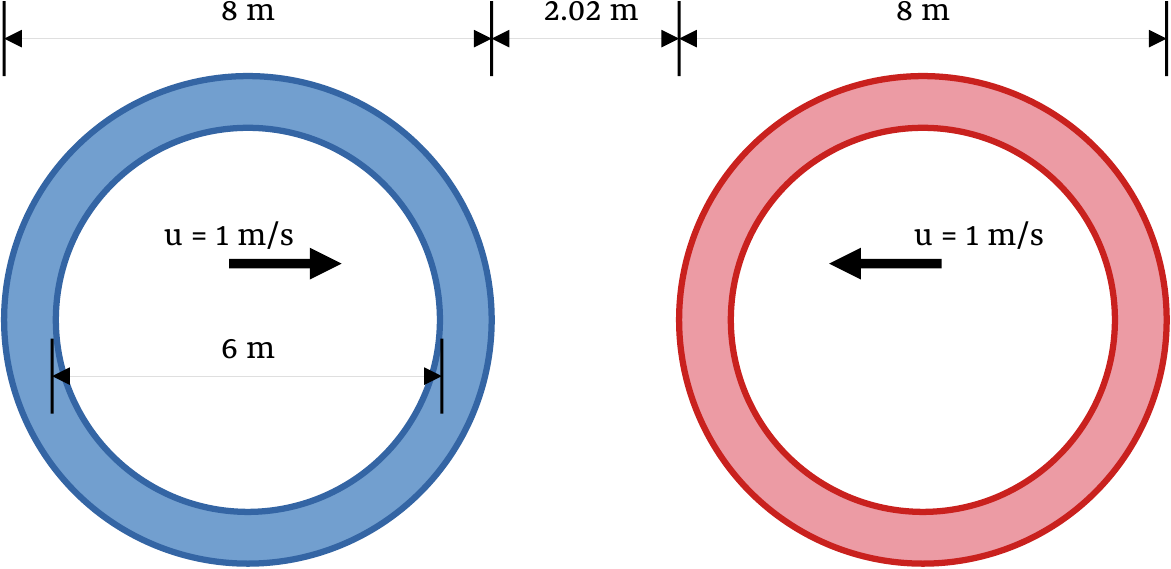}
        \caption{Initial configuration of the collision problem between two Neo-Hookean rings moving toward each other. The initial gap is set to \SI{2.02}{m}.}
        \label{fig:ring-setup}
\end{figure}

The configuration of the colliding-ring benchmark is illustrated in \autoref{fig:ring-setup}. The material parameters are prescribed as $E=50$ kPa, $\nu=0$, and $\rho = 100\,\text{kg}/\text{m}^3$, and a Neo-Hookean constitutive model is used to characterize the nonlinear elastic response throughout the collision. The two rings are initialized with opposing velocities of magnitude $1.0\,\text{m/s}$. All simulations employ APIC transfer and compare linear MPM, CK-MPM, and quadratic B-spline MPM. The linear kernel provides a compact but less smooth baseline, while quadratic B-spline MPM represents a smooth kernel with wider support; CK-MPM is intended to retain smooth transfer behavior with compact support.

Unless otherwise specified, the default spatial and temporal resolutions are set to $\Delta x=\SI{0.125}{m}$ and $\Delta t=\SI{0.001}{s}$, with PPC fixed at $9$ throughout all simulations. In addition, temporal and spatial refinement studies are conducted to investigate the effects of time-step size and grid resolution. For the temporal refinement study, $\Delta x$ is fixed at $\SI{0.125}{m}$, while the time step is varied among $\Delta t=0.0125$, $0.005$, $0.0025$, and $\SI{0.0005}{s}$. For the spatial refinement study, $\Delta t$ is fixed at $\SI{0.001}{s}$, while $\Delta x$ is varied among $0.25$, $0.125$, $0.0625$, and $0.03125\,\mathrm{m}$.

\subsubsection{Stress distribution}

\begin{figure}[htbp]
        \centering
        \includegraphics[width=0.95\textwidth]{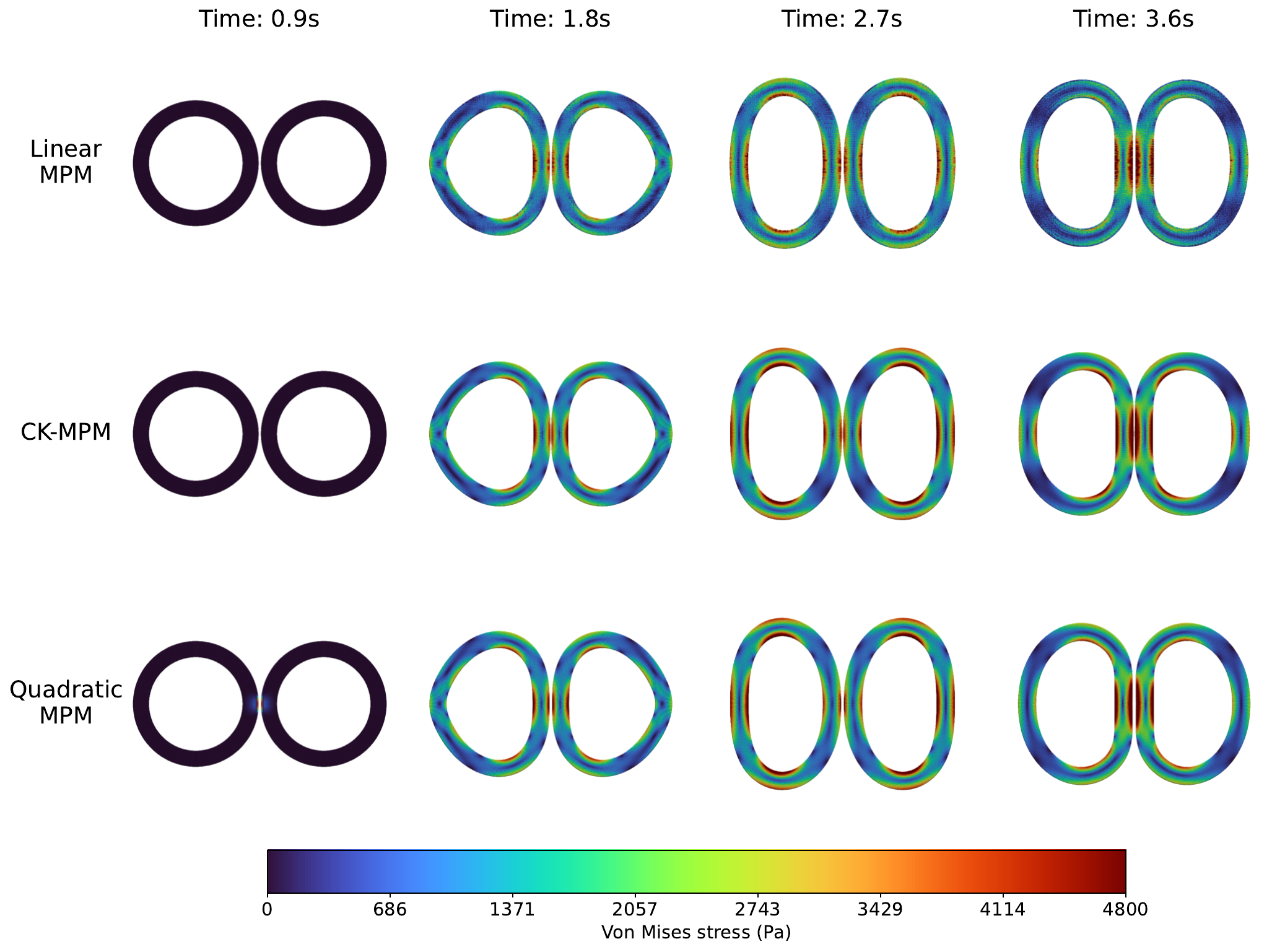}
        \caption{Von Mises stress distributions at selected times. Compared with linear MPM, CK-MPM results in smooth stress distributions without noise. At the same time, it reduces the early-contact artifacts observed in quadratic MPM, thereby more accurately resolving the collision process.}
        \label{fig:ring-stress}
\end{figure}

To investigate the distribution of von Mises stress under different kernel formulations, we compare the stress fields obtained using linear MPM, quadratic B-spline MPM, and CK-MPM. The stress fields of the two colliding rings at $t = \SI{0.9}{s}$, $\SI{1.8}{s}$, $\SI{2.7}{s}$, and $\SI{3.6}{s}$ are presented in \autoref{fig:ring-stress}.

A first observation is that quadratic B-spline MPM exhibits a clear early-contact artifact. By contrast, neither linear MPM nor CK-MPM shows such behavior, owing to their more compact kernel support relative to the quadratic B-spline formulation. Nevertheless, the linear-kernel MPM suffers from substantially stronger stress oscillations under large deformation, as a consequence of the well-known cell-crossing noise associated with discontinuous kernel gradients. In strongly deforming regions, this becomes particularly severe. At time $t = \SI{1.8}{s}$ and $\SI{2.7}{s}$, the stress field predicted by linear MPM deviates significantly from those of the smoother-kernel formulations and loses important local detail, whereas CK-MPM still maintains a stress distribution that is both smooth and physically consistent.

\subsubsection{Energy evolution and temporal refinement}

\begin{figure}[htbp]
        \centering
        \includegraphics[width=0.7\textwidth]{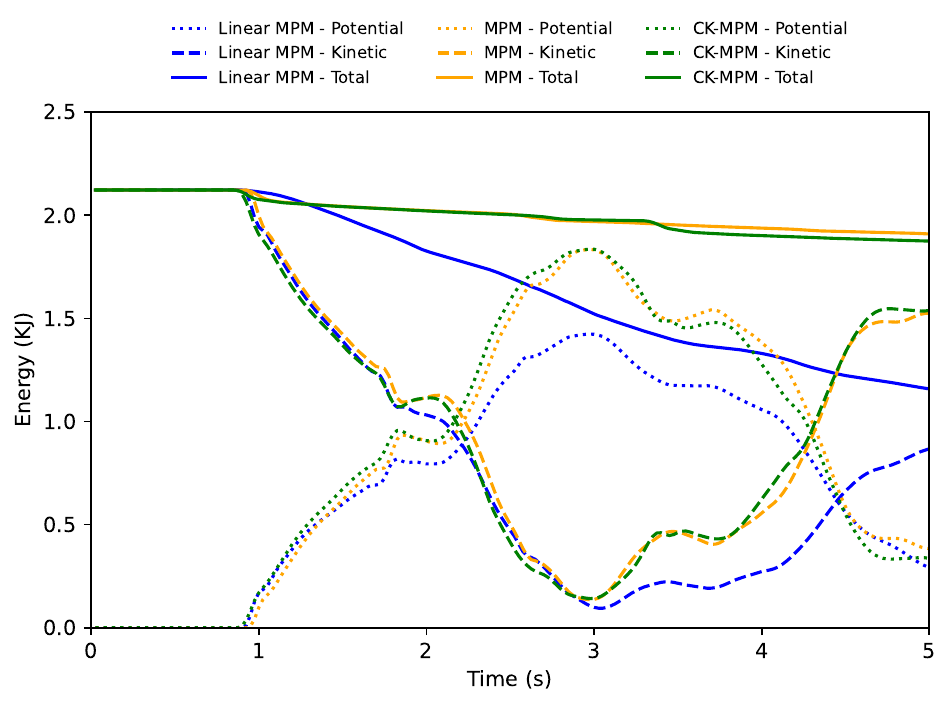}
        \caption{Evolution of the potential, kinetic, and total energies. The energy response of CK-MPM closely follows that of quadratic MPM, indicating that replacing the linear kernel with the compact kernel substantially reduces the excessive energy dissipation observed in linear MPM.}
        \label{fig:ring-energy}
\end{figure}

We next examine the energy conservation characteristics of the three methods. In the absence of artificial numerical dissipation, energy is exchanged only between kinetic energy and elastic strain energy, while the total mechanical energy remains unchanged. The energy histories at $\Delta x=\SI{0.125}{m}$ and $\Delta t=\SI{0.001}{s}$ are shown in \autoref{fig:ring-energy}. CK-MPM closely follows quadratic B-spline MPM in both the kinetic--elastic energy exchange and the total energy evolution. By contrast, linear MPM exhibits a substantially larger total energy drop after impact, indicating greater numerical dissipation in the complete implicit scheme.

\begin{figure}[htbp]
        \centering
        \includegraphics[width=0.7\textwidth]{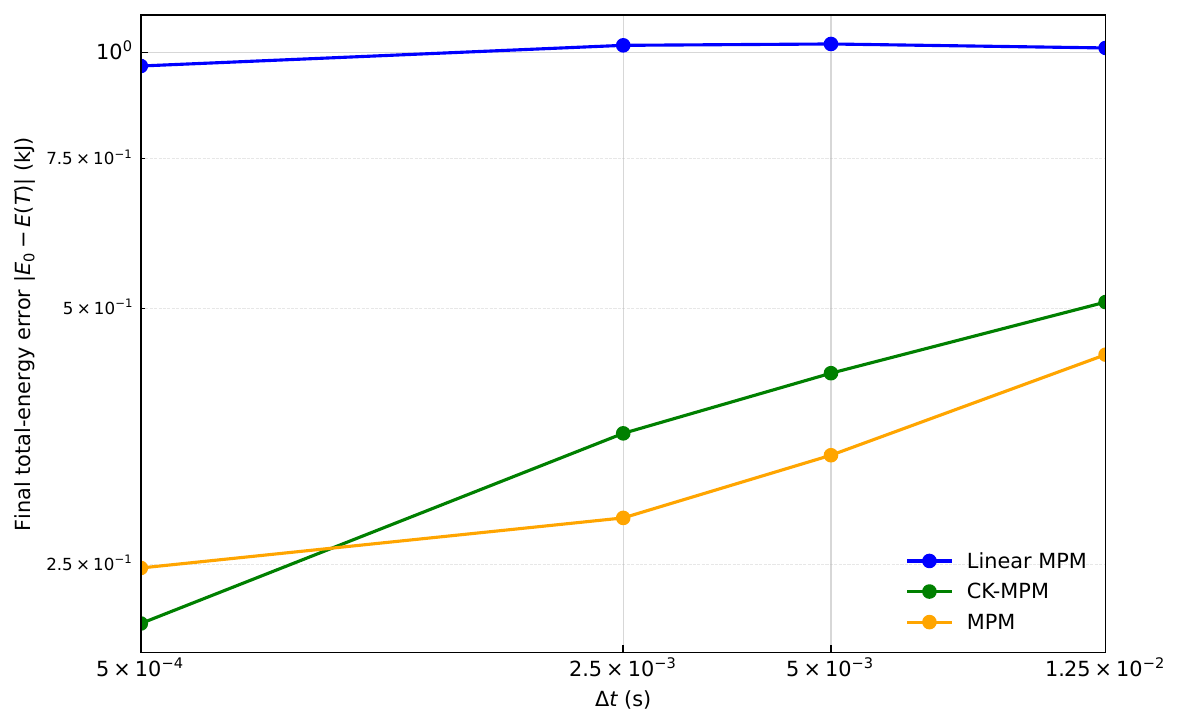}
        \caption{Final total-energy error $|E_0-E(T)|$ at $T=\SI{5}{s}$ under temporal refinement at fixed $\Delta x=\SI{0.125}{m}$. The reference energy $E_0$ is the theoretical initial total energy of the isolated two-ring system.}
        \label{fig:ring-energy-loss-dt}
\end{figure}

For the temporal sweep, the initial total energy is a known conserved reference. At $\Delta x=\SI{0.125}{m}$, its theoretical value is $E_0=\SI{2.122}{kJ}$. We therefore measure the final-time energy error as $|E_0-E(T)|$, where $T=\SI{5}{s}$. \autoref{fig:ring-energy-loss-dt} shows this error over the time step sweep. Reducing $\Delta t$ decreases the energy loss for both CK-MPM and quadratic B-spline MPM, and their curves remain close over the tested range. Linear MPM, in contrast, retains an approximately constant and substantially larger energy loss. These results are consistent with the energy histories: CK-MPM and quadratic B-spline MPM have similar energy-dissipation behavior and both lose markedly less energy than linear MPM.

Numerical energy loss in this benchmark can arise from both backward-Euler integration and particle–grid transfer.  \autoref{fig:ring-energy-loss-dt} shows that reducing the time step at fixed spatial resolution decreases the energy loss for CK-MPM and quadratic B-spline MPM, whereas linear MPM retains a substantially larger loss with weak time-step sensitivity. A smaller time step also increases the number of transfer cycles over the same physical duration, so the observed changes reflect the combined effects of time integration and transfer rather than either contribution alone. Across the tested range, CK-MPM maintains substantially lower energy loss than linear MPM and behavior comparable to quadratic B-spline MPM, supporting its favorable energy response under temporal refinement.

\subsubsection{Angular momentum under spatial and temporal refinement}

Finally, we examine angular momentum conservation in the colliding ring benchmark under both spatial and temporal refinement. The material parameters, initial configuration, opposing velocities, and APIC transfer scheme remain unchanged from the previous experiment. In the absence of external torque, the total angular momentum should remain constant at its initial value, which is zero due to the symmetric initial configuration and opposing velocities of the two identical rings. Following the definition of total particle angular momentum for APIC given by \citet{jiang2017angular}, we account for both the translational and affine contributions. In two dimensions, the out-of-plane component is given by
\begin{equation}
L_z(t)=\sum_p m_p\left[x_p(t)v_{p,y}(t)-y_p(t)v_{p,x}(t)+B_{p,yx}(t)-B_{p,xy}(t)\right],
\label{eq:ring-apic-angular-momentum}
\end{equation}
where $B_{p,yx}$ and $B_{p,xy}$ are the corresponding entries of $\bm{B}_p$, as defined in Section~\ref{sec3.2}. Although the APIC particle--grid transfers conserve this combined angular momentum, the fully implicit backward-Euler time integrator employed here does not preserve it exactly. This experiment therefore evaluates the angular momentum conservation properties of the complete implicit time discretization.

We quantify the deviation from the initial angular momentum using the root-mean-square (RMS) error,
\begin{equation}
e_{\mathrm{RMS}}=\left[\frac{1}{N}\sum_{j=1}^{N}L_z(t_j)^2\right]^{1/2},
\qquad t_j=0.025j\,\mathrm{s},\quad N=200.
\label{eq:ring-angular-momentum-rms}
\end{equation}
Since the reference angular momentum is zero, a relative error is undefined, and the absolute RMS error is used instead. To ensure consistent comparisons across different time-step sizes, angular momentum is sampled at fixed intervals of \SI{0.025}{s}, yielding 200 samples over a total duration of \SI{5}{s}. The RMS errors are thus evaluated at identical physical times for all temporal resolutions.

\begin{figure}[htbp]
    \centering
    \includegraphics[width=0.7\linewidth]{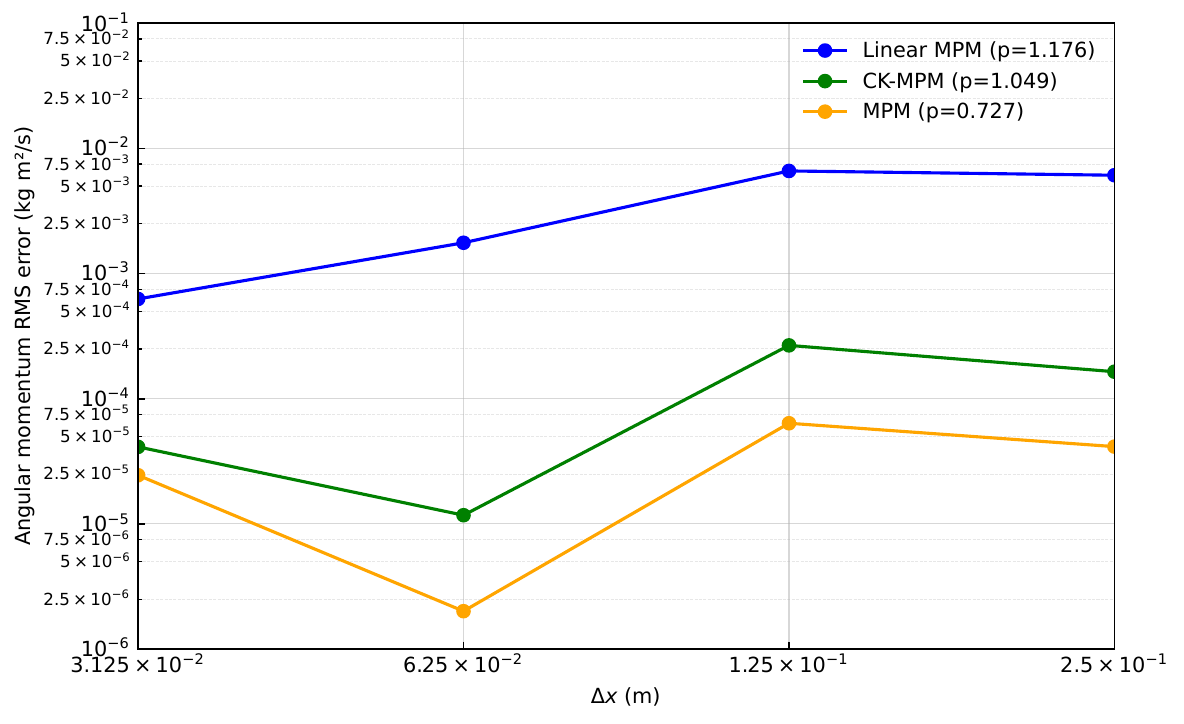}
    \caption{Spatial refinement of the RMS error in total APIC angular momentum, including both translational and affine contributions, for the colliding-ring benchmark. The common comparison spacing $\Delta x$ is reduced while the time step is held fixed at \SI{0.001}{s}.}
    \label{fig:ring-angular-momentum-dx}
\end{figure}

\begin{figure}[htbp]
    \centering
    \includegraphics[width=0.7\linewidth]{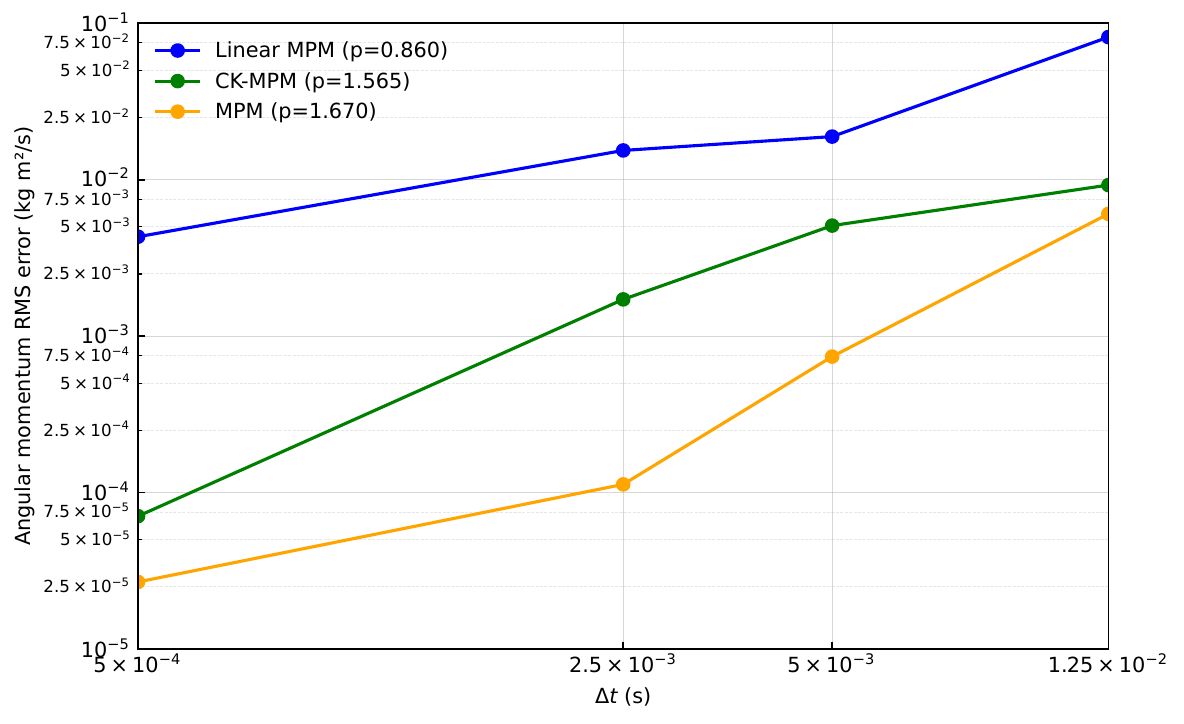}
    \caption{Temporal refinement of the RMS error in total APIC angular momentum, including both translational and affine contributions, for the colliding-ring benchmark. The time step $\Delta t$ is reduced while the comparison spacing is held fixed at \SI{0.125}{m}. All RMS errors are evaluated at the same 200 sampling times over \SI{5}{s}.}
    \label{fig:ring-angular-momentum-dt}
\end{figure}

The spatial and temporal refinement results are shown in \autoref{fig:ring-angular-momentum-dx} and \autoref{fig:ring-angular-momentum-dt}, respectively. For spatial refinement, least-squares fits of $\log e_{\mathrm{RMS}}$ against $\log (\Delta x)$ yield slopes of approximately $1.176$, $1.049$, and $0.727$ for linear MPM, CK-MPM, and quadratic B-spline MPM, respectively. Notably, CK-MPM and quadratic B-spline MPM exhibit similar fluctuations in angular momentum error as the grid is refined, with neither showing a monotonic decrease. Although the fitted slopes indicate an overall reduction in error, this non-monotonic behavior suggests that spatial resolution alone does not govern angular momentum conservation and fully implicit time integrator is a major source of angular momentum error. For temporal refinement, the corresponding fitted slopes are approximately $0.860$, $1.565$, and $1.670$. CK-MPM and quadratic B-spline MPM exhibit similar empirical error reduction rates under temporal refinement, further highlighting the importance of time-step size in controlling angular momentum error. Although CK-MPM yields higher RMS errors than quadratic B-spline MPM, its errors remain substantially lower than those of linear MPM in both refinement studies.

Taken together with the stress and energy results, these findings highlight the balanced performance of implicit CK-MPM. Although CK-MPM exhibits larger angular momentum and energy errors than quadratic B-spline MPM, the two methods show similar error trends, while CK-MPM achieves substantially lower errors than linear MPM. Furthermore, CK-MPM avoids the early-contact artifacts observed with the quadratic B-spline kernel. This balance between accuracy and contact locality is enabled by the combination of a smaller kernel support radius and a smooth kernel function.

\section{Conclusion}
In this work, we developed an implicit formulation of CK-MPM and examined its performance in several benchmark problems representative of computational solid mechanics. The results indicate that the compact-kernel design remains beneficial in the implicit setting, providing an effective balance between transfer smoothness and locality while improving numerical behavior in problems sensitive to diffusion and contact nonlocality.

The numerical studies show that CK-MPM improves upon the excessive stress noise and dissipation associated with linear MPM, while also alleviating the artificial contact-gap and early-contact effects observed in quadratic B-spline MPM. In this sense, CK-MPM offers a practically attractive compromise between highly local but noisy linear kernels and smoother but more nonlocal wide-support kernels.

Overall, the present study suggests that CK-MPM is not limited to its original explicit graphics-oriented context, but can also serve as a viable framework for implicit computational mechanics. The method should not be regarded as a replacement for dedicated contact algorithms, but rather as a kernel-level improvement that can enhance the intrinsic numerical behavior of MPM. Future work will consider its combination with advanced contact treatments, extensions to more challenging three-dimensional and strongly nonlinear problems, and a systematic comparison of the computational efficiency of CK-MPM and standard MPM, including opportunities to improve the efficiency of the dual-grid formulation.

\section{Acknowledgement}
We sincerely thank Zhaofeng Luo, Michael Liu, and Kemeng Huang for valuable discussion. Minchen Li acknowledges Carnegie Mellon University for the Junior Faculty Startup Fund and Genesis AI for the gift funding. Yupeng Jiang acknowledges the financial support from the International Research Training Group (IRTG) 2657, funded by the German Research Foundation (DFG) (Project-ID 433082294).

\newpage

\bibliographystyle{wb_stat}
\bibliography{WileySTAT}

\end{document}